\newcommand{\rcor}[0]{$R_{\rm cr}$}
\newcommand{\rbar}[0]{$R_{\rm bar}$}
\newcommand{\omegabar}[0]{$\Omega_{\rm bar}$}
\newcommand{\sbar}[0]{$S_{\rm bar}$}
\newcommand{\rr}[0]{${\cal R}$}
\newcommand{\vcirc}[0]{$V_{\rm circ}$}

\newcommand{\rmean}[0]{$R_{\rm mean}$}
\newcommand{\rqb}[0]{$R_{\rm Qb}$}

\documentclass{aa}
\usepackage{caption}  

\usepackage[T1]{fontenc}
\usepackage{ae,aecompl}

\usepackage{graphicx}	
\usepackage{amsmath}	
\usepackage{amssymb}	
\usepackage[dvipsnames,usenames,table]{xcolor}
\usepackage{longtable}

\begin{document}

   \title{Bar pattern speeds in CALIFA galaxies}

   \subtitle{III. Solving the puzzle of ultrafast bars}

   \author{Virginia Cuomo\inst{1,2}
          \and
          Yun Hee Lee\inst{3}
          \and
          Chiara Buttitta\inst{2}
          \and
          J. Alfonso L. Aguerri\inst{4,5}
          \and
          Enrico Maria Corsini\inst{2,6}
          \and
          Lorenzo Morelli\inst{1}
          %
          }

   \institute{Instituto de Astronom\'ia y Ciencias Planetarias, Universidad de Atacama, Avenida Copayapu 485, Copiapó, Chile\\
    \email{virginia.cuomo@uda.cl}
         \and
            Dipartimento di Fisica e Astronomia "G. Galilei", Università di Padova, vicolo dell'Osservatorio 3, I-35122 Padova, Italy
        \and
            Korea Astronomy and Space Science Institute, 776 Daedeokdae-ro, Yuseong-gu, 34055 Daejeon, Korea
        \and
            Instituto de Astrof\'isica de Canarias, calle V\'ia Láctea s/n, E-38205 La Laguna, Tenerife, Spain
        \and
            Departamento de Astrof\'isica, Universidad de La Laguna, Avenida Astrof\'isico Francisco Sánchez s/n, E-38206 La Laguna, Tenerife, Spain
        \and
            INAF - Osservatorio Astronomico di Padova, vicolo dell'Osservatorio 2, I-35122 Padova, Italy}

   \date{Received ; accepted }

 
  \abstract
   {More than 10\% of the barred galaxies with a direct measurement of the bar pattern speed host an ultrafast bar. These bars extend well beyond the corotation radius and challenge our understanding of the orbital structure of barred galaxies. Most of them are found in spiral galaxies, rather than in lenticular ones.}
   {We analysed the properties of the ultrafast bars detected in the Calar Alto Legacy Integral Field Spectroscopy Area Survey to investigate whether they are an artefact resulting from an overestimation of the bar radius and/or an underestimation of the corotation radius or a new class of bars, whose orbital structure has not yet been understood.}
   {We revised the available measurements of the bar radius based on ellipse fitting and Fourier analysis and of the bar pattern speed from the Tremaine-Weinberg method. In addition, we measured the bar radius from the analysis of the maps tracing the transverse-to-radial force ratio, which we obtained from the deprojected $i$-band images of the galaxies retrieved from the Sloan Digital Sky Survey.}
   {We found that nearly all the sample galaxies are spirals with an inner ring or pseudo-ring circling the bar and/or strong spiral arms, which hamper the measurement of the bar radius from the ellipse fitting and Fourier analysis. According to these methods, the bar ends overlap the ring or the spiral arms making the adopted bar radius unreliable. On the contrary, the bar radius from the ratio maps are shorter than the corotation radius. This is in agreement with the theoretical predictions and findings of numerical simulations about the extension and stability of the stellar orbits supporting the bars.}
   {We conclude that ultrafast bars are no longer observed when the correct measurement of the bar radius is adopted. Deriving the bar radius in galaxies with rings and strong spiral arms is not straightforward and a solid measurement method based on both photometric and kinematic data is still missing.}


   \maketitle
%

\section{Introduction}

Many disc galaxies, including the Milky Way, host a central bar which contains up to $\sim 30\%$ of the total light \citep{Marinova2007,barazza2008,Aguerri2009,BlandHawthorn2016}. This structure forms from internal or externally-induced dynamical instabilities in a differentially rotating stellar disc and tumbles around the galaxy centre \citep{Sellwood1981,Noguchi1988,Friedli1999,MartinezValpuesta2017}.

Three main properties allow to fully describe a bar: its radius, strength, and pattern speed \citep[e.g.,][]{Aguerri2015}. The bar radius and strength can be derived analysing optical and/or near-infrared images, while the bar pattern speed is a dynamical parameter, which requires kinematic measurements. The bar radius \rbar\ is defined as the length of the bar semi-major axis and measures the extension of the stellar orbits supporting the bar \citep{contopoulos1980,Contopoulos1981}. However, bars do not usually present sharp edges and are often associated with other components (such as rings and/or spiral arms), so it is difficult to properly identify the bar boundaries \citep{Aguerri2009}. In turn, the presence of a large bulge complicates the analysis \citep{aguerri2005}. Several methods have been developed to derive \rbar, but each of them suffers from some limitations \citep[see e.g.,][for a discussion]{Corsini2011}. In order to overcome the problems related to the choice of a single method, usually \rbar\ is estimated combining the results of different independent methods \citep{Corsini2003,Guo2019,Cuomo2020}. The bar strength \sbar\ describes the contribution given by the non-axisymmetric mass density of the bar to the galaxy gravitational potential \citep{buta01} and it can be used to distinguish between strong and weak bars \citep{Cuomo2019b}. The results based on the strength of bars are quite controversial because \sbar\ strongly depends on the method used to measure it \citep{Lee2020}. The bar pattern speed \omegabar\ is the angular frequency of the bar tumbling around the galaxy centre, and it determines the so-called corotation radius \rcor, which is the radius where the angular velocity of the disc \vcirc\ is equal to \omegabar. Bars can be classified according to the rotation rate parameter, defined as ${\cal R}=R_{\rm cr}/R_{\rm bar}$ \citep{Elmegreen1996bis}. 

Bars are mainly supported by elongated stellar orbits called $x_1$ orbits, which are stable within \rcor. In contrast, $x_1$ orbits outside \rcor\ are arranged perpendicular to the bar major axis and do not sustain the bar structure \citep{contopoulos1980,Contopoulos1981,Vasiliev2015}. Current dynamical arguments show that bars can not have ${\cal R} < 1.0$. On the other hand, bars with ${\cal R} \geq 1.0$ are usually classified either as fast ($1.0 \leq {\cal R} \leq 1.4$) or slow (${\cal R} > 1.4$). Fast bars end close to corotation and rotate as fast as they can, while slow bars are shorter and rotate more slowly. The separation between long/fast and short/slow bars does not imply a specific value of \omegabar, while the dividing value at 1.4 is commonly used \citep{Athanassoula1992,Debattista2000}. In fact, the definition of \rr\ was introduced by \citet{Elmegreen1996bis} to show how bars usually end inside corotation, possibly between the ultra-harmonic 4:1 resonance and corotation. However, the bar-spiral transition does not occur at corotation and the spiral arms can extend for a significant distance inside corotation, since spiral density waves can propagate between the inner and the outer Lindblad resonances \citep{Adams1989,Bertin1996}.

As a consequence of the angular momentum exchange within the galaxy and of the dynamical friction exerted on the bar by the dark matter (DM) halo, \omegabar\ is expected to decrease with time during the evolution \citep{weinberg1985,little1991,Debattista1998,oneill2003,villavargas2010,Athanassoula2013}. The slow down of the bar is stronger if a massive and centrally-concentrated DM halo is present, because there is more mass ready to absorb angular momentum near the resonances and dynamical friction is more efficient (see \citealt{Athanassoula2014} and \citealt{Sellwood2014}, for further discussion). This implies that galaxies hosting fast bars should be embedded in DM halos with a low central density, such as those required in the maximum disc hypothesis \citep{Debattista2000,Palunas2000,Fuchs2001,Starkman2018}. Nevertheless, other galaxy properties may influence the angular momentum exchange within the galaxy, such as the halo triaxiality, presence of gas, and disc velocity dispersion \citep{Athanassoula2003,Athanassoula2013}. The measurement of \rr\ may help to disentangle the DM distribution and to investigate the secular evolution of barred galaxies.

There are several methods to recover \omegabar\ and consequently \rr, but most of them require some modelling \citep{Kormendy1979,Athanassoula1992,lindblad1996,puerari1997,Aguerri2000,Zhang2007,Rautiainen2008,font2011}. \citet{Tremaine1984} developed a technique to recover \omegabar, hereafter called TW method, which does not require any dynamical model. When the bar is characterised by an exact solid-body rotation, \omegabar\ is directly determined from observable quantities measured for a tracer population of stars or gas, which only has to satisfy the continuity equation. The method requires to measure the surface brightness and radial velocity of the tracer along apertures located parallel to the line-of-nodes. When both positions and velocities are measured with respect to the galaxy centre, then the luminosity-weighted mean velocity divided by the luminosity-weighted mean position gives $\Omega_{\rm bar}\sin i$, where $i$ is the disc inclination.

More than 100 galaxies have been analysed so far with the TW method, each providing an estimate for \rr\ \citep[see][for a review]{Cuomo2020}. Neglecting measurements with large uncertainties, $\sim90\%$ of the bars are consistent with the fast regime at $95\%$ confident level. These galaxies should host little DM in their central regions, in agreement with the findings based on the study of rotation curves for unbarred galaxies \citep{Debattista2000,Starkman2018}.

Despite theoretical predictions, $\sim10\%$ of the galaxies with \omegabar\ measured with the TW method shows ${\cal R} < 1.0$ at $95\%$ confident level: these bars are termed as `ultrafast' \citep{Buta2009}. According to their Calar Alto Legacy Integral Field Spectroscopy Area Survey \citep[CALIFA;][]{sanchez2012,Walcher2014,FalconBarroso2017} morphological classification, all of them are found in late-type barred galaxies with Hubble stage $T$ between 2 and 7, except for the lenticular galaxy NGC~2880 \citep{Cuomo2020}. Whether ultrafast bars are the consequence of an erroneous application of the TW method or a special class of bars, which do not obey to the predictions of theory and numerical simulations (see \citealt{Aguerri2015} and \citealt{Guo2019}, for a discussion). However, a non-negligible fraction of ultrafast bars was also observed while applying other methods \citep{Buta2009,Buta2017}. \citet{Garma-Oehmichen2020} analysed the different sources of error in the TW method and quantified how much they are affecting the final measurement of \omegabar\ in a sample of 15 galaxies. The dominant sources of error are the identification of the disc position angle PA and the length of the apertures along which to measure the position and velocity of the tracer, while centring errors and a degraded point-spread-function result in a small or negligible effect. \cite{Garma-Oehmichen2020} did not observe a significant correlation between the error sources, but they stressed the importance of the correct error treatment. In fact, they claimed that a large fraction of ultrafast bars may be the result of an erroneous treatment of the errors together with low spatial resolution data. New studies are needed to eventually exclude that these results are flawed because of an improper application of the TW method and/or to investigate if some information about the nature of ultrafast bars is still missing. 

Recent studies pointed out that spiral arms may affect the measurement of \rbar\ \citep{Petersen2019,Hilmi2020}. In particular, \citet{Hilmi2020} showed with their simulations that the measurement of \rbar\ dramatically fluctuates on a dynamical timescale depending on the strength of the spiral structure and on the measurement threshold. In this paper, we aim to test whether the measurements of \rbar\ adopted in literature may be biased by the presence of the spiral arms or other components, which caused the extremely low values of \rr. 

We organise the paper as follows: we introduce the sample of galaxies and their properties as available in literature in Sect.~\ref{sec:method_sample}. We present our analysis in Sect.~\ref{sec:results}. We discuss our results and present our conclusions in Sects.~\ref{sec:discussion} and \ref{sec:conclusions}, respectively. We adopt as cosmological parameters, $\Omega_m = 0.286,~\Omega_\Lambda = 0.714$, and $H_0 = 69.3$ km s$^{-1}$ Mpc$^{-1}$ \citep{Hinshaw2013}.

\section{Sample selection and available data}
\label{sec:method_sample}

\begin{figure*}
    \centering
    \includegraphics[scale=0.56]{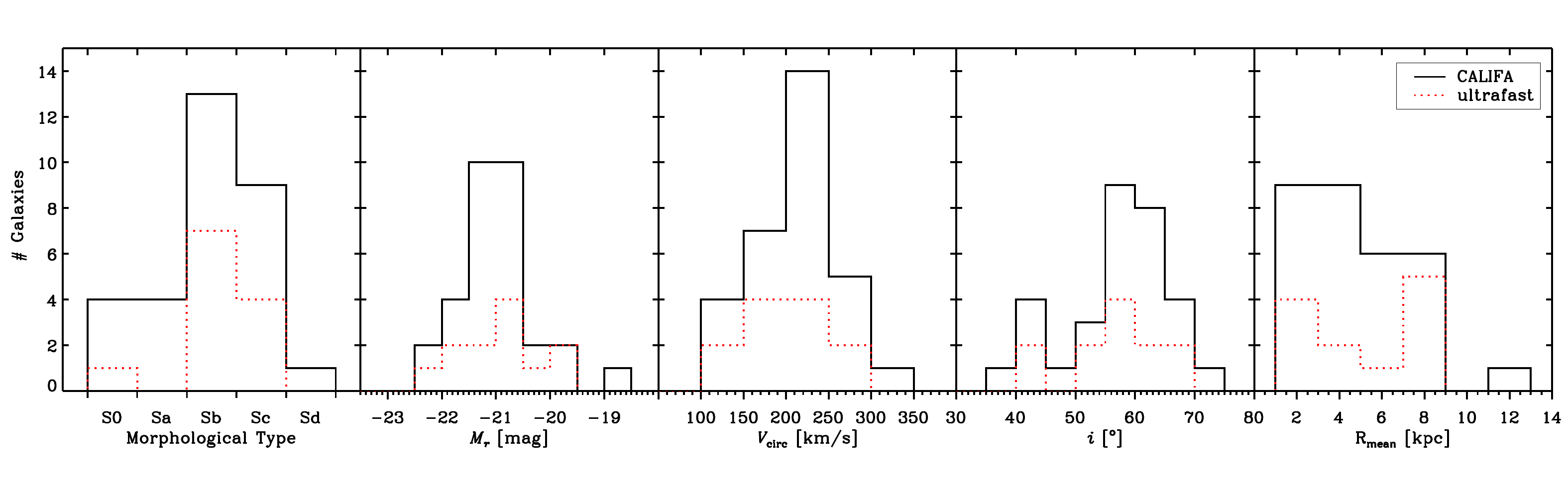}
    \caption{Distribution of the morphological types, absolute SDSS $r$-band magnitudes, circular velocities, disc inclinations and bar radii of the initial sample of 31 CALIFA galaxies analysed with the TW method by \cite{Aguerri2015} and \cite{Cuomo2019b} (black solid line) and of the subsample of 12 galaxies hosting an ultrafast bar discussed in this paper (red dotted line).}
    \label{fig:histo_gal}
\end{figure*}

We considered the 31 barred galaxies studied by \citet{Aguerri2015} and \citet{Cuomo2019b} for which a direct TW measurement of \omegabar\ was obtained by analysing the stellar kinematics from CALIFA. \citet{Aguerri2015} selected 15 strongly-barred galaxies, while \citet{Cuomo2019b} focused their attention onto 16 weakly-barred galaxies. The resulting sample spans a wide range of morphological types (SB0\,--\,SBd), redshifts (0.005\,--\,0.02), and absolute SDSS $r$-band magnitudes ($-18.5$\,--\,$-22.0$ mag). These distributions are similar to those of the total sample of bright barred galaxies targeted by the CALIFA survey. Twelve galaxies measured by \citet{Aguerri2015} and \citet{Cuomo2019b} turned out to host an ultrafast bar, according to their values of \rr\ and corresponding errors. Figure~\ref{fig:histo_gal} shows the distributions of morphological types, absolute SDSS $r$-band magnitudes, circular velocities, disc inclinations, and bar radii of the initial sample of 31 galaxies compared to those of the subsample of 12 galaxies hosting an ultrafast bar. We ran a Kolmogorov–Smirnov test with the \textsc{idl} procedure \textsc{kstwo} and verified there are no statistical differences at a very high confidence level ($>95\%$) between the distributions of properties of the two samples. All the ultrafast bars except for one are found in SBab\,--\,SBc spiral galaxies.

The TW method was applied to recover \omegabar\ (and consequently \rr) by using the following straightforward equation

\begin{equation}
\langle V \rangle = \langle X \rangle ~\Omega_{\rm bar}~\sin i
\end{equation}

\noindent which requires to know the disc inclination $i$,

\begin{equation}
\langle X\rangle=\frac{\int X \Sigma {\rm d}X}{\int \Sigma {\rm d}X} ,\;\;{\rm and}\;\; \langle V\rangle=\frac{\int V_{\rm los}\Sigma {\rm d}X}{\int \Sigma {\rm d}X}
\label{eq.2}
\end{equation}

\noindent which are defined as the luminosity-weighted average of position $X$ and line-of-sight (LOS) velocity $V_{\rm los}$ of the stars, respectively, where $\Sigma$ is the surface brightness. These integrals are measured along several pseudo-slits. One is centred on the galaxy centre and the others have an offset, but all are aligned with the disc major axis, which requires the determination of the disc PA. 

As described in \citet{Aguerri2015} and \citet{Cuomo2019b}, the disc parameters ($i$ and PA) are obtained by analysing the outer isophotes of the galaxy and the resulting radial profiles of the ellipticity $\epsilon$ and PA. The galaxy isophotes were modelled using the \textsc{ellipse} routine of the \textsc{iraf\footnote{Image Reduction and Analysis Facility is distributed by the National Optical Astronomy Observatories,
which are operated by the Association of Universities for Research in Astronomy under cooperative agreement with the National Science Foundation.}} package \citep{Jedrzejewski1987}. 

The slope of the straight line defined by the integrals in Eq.~\ref{eq.2} gives $\Omega_{\rm bar}\sin i$. In practice, the luminosity-weighted photometric and kinematic integrals are obtained by collapsing the spectroscopic datacube along the wavelength and spatial directions of each pseudo-slit, respectively. For the sample galaxies, we adopted the values of \omegabar\ obtained in this way by \citet[][see col. 4 in their Table 3]{Aguerri2015} and \citet[][see col. 3 in their Table 2]{Cuomo2019b}. As an alternative, the kinematic integrals can be directly obtained from the stellar velocity field. Moreover, either a map of the stellar surface brightness \citep{Aguerri2015,Guo2019,Garma-Oehmichen2020} or stellar surface mass density \citep{Aguerri2015,Guo2019,Williams2021} can be used as a weight in the definition of the integrals. However, the mass and light distributions often do not match well, particularly in the presence of ongoing star formation, as is usually the case in spiral galaxies.

The radius of \rbar\ is usually obtained as the mean value of various measurements retrieved using several methods \citep{Corsini2003,Cuomo2019,Guo2019}. Indeed, three different methods were adopted in \citet{Aguerri2015} and \citet{Cuomo2019b} to recover \rbar\ based on the photometric analysis of Sloan Digital Sky Survey images \citep[SDSS;][]{Abazajian2009}. Hereafter, we refer to this estimate as \rmean. 

The first two methods are based on the study of the radial profile of $\epsilon$ and PA of the ellipses which best fit the galaxy isophotes \citep{MenendezDelmestre2007,Aguerri2009}. The bar radius corresponds to the position of the peak in the ellipticity profile, $R_{\rm \epsilon,peak}$, or to the position where the PA changes by $\Delta{\rm PA}=5\degr$ from the PA of the ellipse with the maximum $\epsilon$ value, $R_{\rm PA}$. Undisturbed barred galaxies usually show a local peak and a sudden outward decrease of $\epsilon$ to a minimum value ($\Delta \epsilon \geq 0.08$), which corresponds to the region of the disc where the isophotes become circular in the face-on case \citep{Wozniak1991,Athanassoula1992,Aguerri2000}. On the other hand, the radial profile of PA is constant in the bar region ($\Delta {\rm PA} \leq20 \degr$) \citep{Wozniak1991,Aguerri2000}. These peculiarities are due to the shape and orientation of the stellar orbits of the bar \citep{Contopoulos1989,Athanassoula1992}. These methods however require calibration on mock galaxy images in order to set the maximum variation of each isophotal parameter which constrains the corresponding \rbar. 
The third adopted approach to measure \rbar\ consists in the Fourier decomposition of the galaxy light and in the analysis of the bar/interbar intensity ratio and provides $R_{\rm Fourier}$ \citep{Ohta1990,Aguerri2000}. The deprojected azimuthal radial profile of the luminosity of the galaxy $I(r,\phi)$ can be described with a Fourier series

\begin{equation}
    I(r,\phi)=\frac{A_0(r)}{2}+\sum_{m=1}^{\infty}[A_m(r)\cos({m\phi})+B_m(r)\sin({m\phi})]
\end{equation}

\noindent where the Fourier components are defined by

\begin{eqnarray}
A_m(r)=\frac{1}{\pi}\int_0^{2\pi} I(r,\phi)\cos({m\phi}) {\rm d}\phi \\ B_m(r)=\frac{1}{\pi}\int_0^{2\pi} I(r,\phi)\sin({m\phi}) {\rm d}\phi,
\end{eqnarray}

\noindent the Fourier amplitude of the $m$-th component is defined as

\begin{equation}
I_m(r)=\begin{cases} 
A_0(r)/2  & \mbox{if }m = 0 \\ \sqrt{A_m^2(r)+B_m^2(r)} & \mbox{if }m \neq 0.
\end{cases}
\end{equation}

\noindent In the bar region, the relative amplitudes $I_m/I_0$ of the even Fourier components ($m=2,4,6,...$) are larger than the odd ones ($m=1,3,5,...$), and the dominant one is the $m=2$. Through this analysis, $R_{\rm Fourier}$ can be recovered from the luminosity contrast between the bar and interbar intensity as a function of radial distance \citep{Aguerri2000}. The intensity of the bar is defined as $I_{\rm bar}=I_0+I_2+I_4+I_6$, while that of the interbar is defined as $I_{\rm ibar}=I_0-I_2+I_4-I_6$. The bar region is where the bar/interbar intensity ratio $I_{\rm bar}/I_{\rm ibar} > 0.5 \times [\max (I_{\rm bar}/I_{\rm ibar})-\min (I_{\rm bar}/I_{\rm ibar})]+\min (I_{\rm bar}/I_{\rm ibar})$ and $R_{\rm Fourier}$ corresponds to the full width at half maximum (FWHM) of the curve given by $I_{\rm bar}/I_{\rm ibar}$ as a function of radius. This method was tested with $N$-body simulations resulting to provide an error of 4\% on the corresponding \rbar\ \citep{Athanassoula2002}. However, it can be hampered by the presence of a non-axisymmetric disc and strong spiral arms. 

The bar rotation rate \rr\ is obtained as the ratio between \rcor\ and \rbar. In turn, \rcor\ is given by the ratio between \omegabar\ and \vcirc, which is usually estimated with an asymmetric drift correction of the observed stellar streaming velocities \citep{Merrifield1995,Debattista2002,Aguerri2015,Cuomo2019b}.

\begin{figure*}[!h]
    \centering
    \includegraphics[scale=0.9]{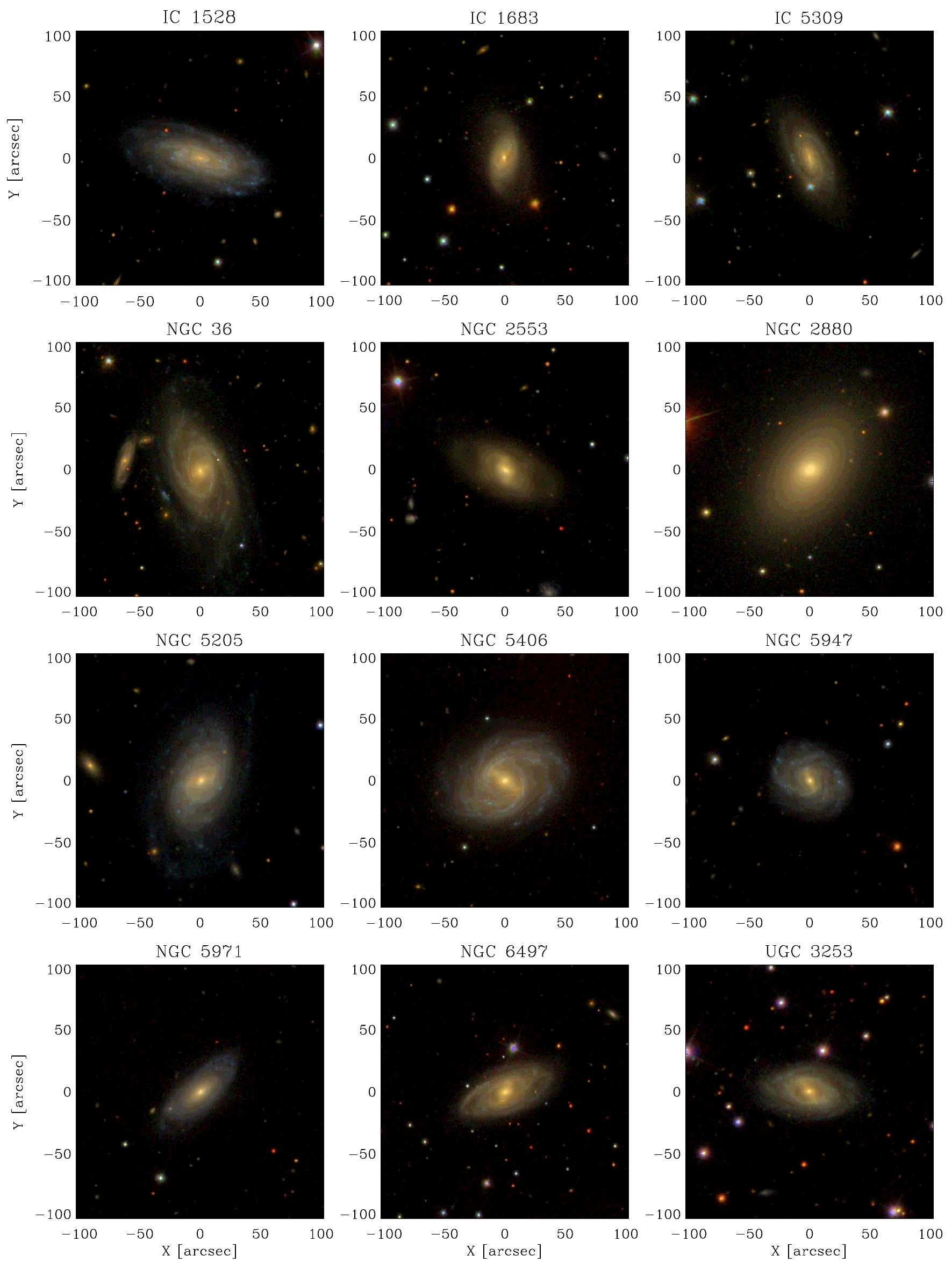}
    \caption{SDSS color images of the sample galaxies. The field of view is $100\times100$ arcsec$^{2}$ and it is oriented with north up and east left.}
    \label{fig:sample}
\end{figure*}

In our analysis, the bar is considered to be ultrafast not only if the corresponding \rr\ is lower that 1.0 at 95\% confident level (as done in \citealt{Cuomo2020}), but when the sum between \rr\ and its upper error is lower than 1.0. This choice allows us to build a better defined sample of possible ultrafast bars. The corresponding sample consists of 12 galaxies with properties presented in Table~\ref{tab:properties} and corresponding SDSS color images given in Fig.~\ref{fig:sample}.

\begin{table*}[]
\caption[Properties of the galaxies.]{Properties of the sample barred galaxies hosting an ultrafast bar.}
    \centering
    \renewcommand{\tabcolsep}{0.04cm}
    \begin{tabular}{cccccccccccc}
    \hline\hline
Galaxy & Morph. Type & Hubble Type & $z$ & $M_r$ & $i$ & PA & $R_{\rm mean}$ & \omegabar\ & \rcor\ & \rr\ & Ref.\\
 & & & & [mag] & [\degr] & [\degr] & [arcsec] & [km s$^{-1}$ arcsec$^{-1}$] & [arcsec] &  &  \\
 (1) & (2) & (3) & (4) & (5) & (6) & (7) & (8) & (9) & (10) & (11) & (12)\\ 
\hline
IC~1528 & SABbc & SAB(rs)b & 0.013 & $-20.57$ & 66.7 & 72.7 & 8.89$^{+2.73}_{-2.93}$ & $21.0\pm3.8$ & $6.74\pm2.11$ & 0.76$^{+0.14}_{-0.22}$ & 2\\
IC~1683 & SABb & S? & 0.016 & $-20.73$ & 54.3 & 13.0 & 27.39$^{+1.93}_{-2.03}$ & $9.7\pm0.4$ & $19.72\pm8.47$ & $0.72\pm0.21$ & 2\\
IC~5309 & SABc & Sb & 0.014 & $-19.99$ & 60.0 & 26.7 & 7.39$^{+3.32}_{-1.87}$ & $24.3\pm7.5$ &  $4.67\pm3.77$ & 0.63$^{+0.35}_{-0.45}$ & 2\\
NGC~36 & SBb & SAB(rs)b & 0.020 & $-21.86$ & 57.2 & 23.4 & 20.19$^{+5.09}_{-4.51}$ & $17.4\pm5.2$ & 12.60$^{+5.39}_{-3.91}$ & 0.6$^{+0.3}_{-0.2}$ & 1 \\
NGC~2553 & SABab & S? & 0.016 & $-21.23$ & 54.6 & 67.0 & 22.16$^{+5.97}_{-5.22}$ & $23.6\pm1.7$ & $11.40\pm2.63$ & 0.51$^{+0.08}_{-0.11}$ & 2 \\
NGC~2880 & EAB7 & SB0$^-$ & 0.005 & $-20.34$ & 56.7 & 144.6 & 12.77$^{+6.09}_{-3.60}$ & $22.2\pm1.3$ & $9.43\pm3.09$ & 0.74$^{+0.15}_{-0.19}$ & 2\\
NGC~5205 & SBbc & S? & 0.006 & $-19.65$ & 50.0 & 170.1 & 17.69$^{+2.83}_{-2.07}$ & $15.1\pm2.8$ & 11.34$^{+2.99}_{-2.53}$ & 0.7$^{+0.2}_{-0.1}$ & 1 \\
NGC~5406 & SBb & SAB(rs)bc & 0.018 & $-22.25$ & 44.9 & 111.8 & 21.00$^{+1.09}_{-2.10}$ & $22.8\pm8.0$ & 11.01$^{+4.80}_{-3.00}$ & 0.5$^{+0.2}_{-0.1}$ & 1 \\
NGC~5947 & SBbc & SBbc & 0.020 & $-21.28$ & 44.6 & 72.5 & 10.91$^{+1.29}_{-1.60}$ & $31.7\pm4.2$ & 5.79$^{+2.39}_{-2.30}$ & $0.5\pm0.2$ & 1 \\
NGC~5971 & SABb & Sa & 0.011 & $-20.57$ & 69.0 & 132.0 & 23.85$^{+20.10}_{-10.91}$ & $16.9\pm4.3$ & $13.37\pm6.44$ & 0.56$^{+0.15}_{-0.32}$ & 2 \\
NGC~6497 & SBab & SB(r)b & 0.010 & $-21.72$ & 60.9 & 112.0 & 14.70$^{+2.09}_{-1.29}$ & $42.7\pm7.4$ & 5.49$^{+2.09}_{-1.60}$ & $0.3\pm0.1$ & 1\\
UGC~3253 & SBb & SB(r)b & 0.014 & $-20.65$ & 56.8 & 92.0 & 15.81$^{+1.29}_{-2.20}$ & $15.5\pm3.1$ & 11.89$^{+3.18}_{-2.69}$ & 0.7$^{+0.2}_{-0.2}$ & 1\\
\hline
    \end{tabular}
    \\
\tablefoot{(1) Galaxy name. (2) Morphological classification from CALIFA \citep{Walcher2014}. (3) Hubble type from \citet[][hereafter RC3]{deVaucouleurs1991}. (4) Redshift from SDSS-DR14 \citep{Abolfathi2018}. (5) Absolute SDSS $r$-band magnitude obtained from the model $r$-band apparent magnitude $m_r$ provided by the SDSS-DR14 and the galaxy distance from NED\footnote{The NASA/IPAC Extragalactic Database is available at \url{https://ned.ipac.caltech.edu/}} as obtained from the radial velocity with respect to the cosmic microwave background reference frame. (6) Disc inclination. (7) Disc position angle. (8) Deprojected bar radius. (9) Bar pattern speed. (10) Bar corotation radius. (11) Bar rotation rate. (12) Source of disc and bar properties: 1 = \citet{Aguerri2015}, 2 = \citet{Cuomo2019b}.} 
    \label{tab:properties}
\end{table*}

\section{Revising the determination of the bar rotation rate}
\label{sec:results}

The determination of \rr\ involves two different quantities which characterise the bar: its radius and the corotation radius. For our sample, the determination of \rcor\ was obtained with the TW method, so the corresponding sources of error have to be considered. On the other hand, \rbar\ was calculated using different methods based on photometry, which may be affected by their own limitations. 

\subsection{Sources of error for the TW analysis}
\label{sec:pa_inc}

The successful application of the TW method requires the disc to have an intermediate $i$ and the bar to be located at an intermediate PA with respect to the disc major and minor axes \citep{Corsini2011}. Moreover, recovering bar PA and $\epsilon$ from ellipse fitting can be very difficult when the galaxy is very inclined \citep{Comeron2014}. All the sample galaxies were selected to have an inclination $20\degr \lesssim i \lesssim 70 \degr$ and a PA difference between bar and disc major axis $10\degr \lesssim \Delta {\rm PA} \lesssim 80 \degr$. 

\cite{Cuomo2020} explored the relative errors on \omegabar, \rbar, and \rcor\ for all the galaxies with available TW measurements as a function of the disc $i$ and bar $\Delta {\rm PA}$, respectively (see their Figs. 2 and 3). They did not observe any significant trend and excluded any selection bias on the entire galaxy sample. This result is true for the subsample of galaxies hosting an ultrafast bar as well.

Moreover, the right identification of the disc parameters (especially the PA along which to locate the pseudo-slits) is crucial for the TW method \citep{Debattista2003,Zou2019}. All the galaxies were carefully selected to be suitable for the TW analysis, because they present a constant behaviour of the PA radial profile in the disc region. Moreover, the sources of error associated to the TW method, i.e., uncertainties in the identification of the disc PA and galaxy centre, and in the choice of the length of the pseudo-slits \citep{Corsini2011,Garma-Oehmichen2020}, were fully taken into account in the application of the TW method and propagated in the error estimate of \omegabar.

\subsection{Considerations about galaxy morphology}

In the early application of the TW method, mainly lenticular galaxies were targeted. This was done to face the strict requests of the method, the limitations due to the long-slit spectroscopy, and to avoid morphological peculiarities or multiple pattern speeds which may be associated to spiral arms. After integral-field spectroscopy became largely used and more theoretical studies about the impact of multiple spiral arms on the TW analysis were available \citep{Debattista2002,Meidt2008}, spiral galaxies were measured and a non-negligible fraction of ultrafast bars were found in these galaxies. Indeed, in the initial sample of 31 CALIFA galaxies, 11 out of 27 SBa\,--\,SBd galaxies ($41\%$) host an ultrafast bar, while this is the case only for one out of 4 SB0s ($25\%$). In fact, the presence of spiral arms and other structures makes it more difficult to correctly determine the disc parameters and TW integrals \citep{Corsini2011,Cuomo2020}. Nevertheless, the applicability of the TW method to spiral galaxies was tested and demonstrated by both theoretical \citep{gerssen2007,Zou2019} and observational studies involving high quality integral-field data \citep{Aguerri2015,Guo2019,Garma-Oehmichen2020, Williams2021}. In particular, \citet{gerssen2007} explored the possible influence of the presence of dust on the reliable application of the TW method. In addition, \citet{Aguerri2015} used the distribution of the mass as weight for the integrals, and they found that the results are compatible with the case of light as weight. In particular, they discussed the cases of NGC~36, NGC~5205, and NGC~6947, excluding that the presence of dust may lead to a value of ${\cal R}<1.0$. 

We carefully performed a visual inspection of the color images of the galaxies showed in Fig.~\ref{fig:sample} and inspected the SDSS $g-i$ color images before and after their deprojection. We concluded that most of the galaxy in our sample (7 out of 12, corresponding to $\sim60\%$) host inner rings or pseudo-rings around the bar. Three of them present an outer ring or pseudo-ring too. These structures are often associated with a pronounced light deficit around the bar inside the inner ring, giving rise to a typical "$\theta$" shape, called `dark gap' \citep{Kim2016,James2016,Buta2017}. This is the case for the galaxies: NGC~36, NGC~2553, NGC~5205, NGC~5406, NGC~5947, NGC~6497, and UGC~3253 (Fig.~\ref{fig:sample}). We carefully analysed those features using the prescriptions of \citet{Buta2015}, \citet{Buta2017}, and \citet{Bittner2020}. Multiple spiral arms are clearly visible in most of the sample galaxies except for IC~1683 and NGC~2553, which host a two-armed structure and NGC~2880 which does not host any spiral arm and it is a lenticular galaxy. Flocculent spiral arms are clearly visible in IC~1528, NGC~5406 and NGC~5947 while grand-design ones are visible in IC~5309 and NGC~36. We concluded that our sample galaxies host spiral arms with various geometrical properties, level of symmetry and amplitude, spanning from flocculent, to multi-armed and grand-design morphologies. The results of our morphological analysis are presented in Sect.~\ref{app1} for the entire galaxy sample.

\subsection{Errors in bar radius}

The adopted values of \rbar\ for the sample galaxies, \rmean, correspond to the mean result obtained using three different methods based on photometry. However, several issues may lead to the wrong determination of \rbar. In particular, a late type morphology which includes the presence of strong spiral arms or other structures may hamper the right determination of \rbar\ \citep{Petersen2019,Hilmi2020}. 

In order to obtain an independent method to recover \rbar, we applied the method proposed by \cite{Lee2020} based on the analysis of the maps tracing the transverse-to-radial force ratio $Q_T(r, \phi)$ of the galaxy \citep{Sanders1980,Combes1981}. While all the other methods discussed in Sect.~\ref{sec:method_sample} are based on the study of the light distribution in galaxies, the approach proposed by \cite{Lee2020} involves 
the calculation of the gravitational potential of the galaxy expected from the light distribution. Despite this method is based on photometry too, it allows to obtain an independent \rbar\ estimate, which we call \rqb\ hereafter. In particular, it allows to disentangle the radius corresponding to the maximum strength of the bar from that of the spiral arms and/or rings by investigating the azimuthal profile according to the radius and to test whether \rmean\ truly matches the bar region. At the same time, it is obtained an alternative estimate of \sbar, hereafter called $Q_{\rm b}$. 

After deprojecting the SDSS $i$-band image of the galaxy, the Poisson equation is solved assuming a constant mass-to-light ratio. The gravitational potential is obtained from the Poisson equation by the convolution of the three-dimensional mass density $\rho ({\bf x'})$ and $1/|{\bf x} - \bf x'|$ \citep{Quillen1994,Buta2001}

\begin{equation}
\Psi ({\bf x})=-G \int \frac{\rho ({\bf x'}){\rm d}^3 {\bf x'}}{|{\bf x} - \bf x'|}.
\end{equation}

\noindent The three-dimensional mass density can be written as $\rho ({\bf x'})=\Sigma (x,y) \rho{_z}(z)$, where $\Sigma (x,y)$ is the mass surface density in the plane of the galaxy and $\rho{_z}(z)$ is the normalized vertical density distribution assumed to follow an exponential profile. In the galaxy plane $z=0$, the two-dimensional potential can be defined in polar coordinates $\Phi(r,\phi)\equiv \Psi (x,y,z=0)$.

The mean radial force $\langle F_{\rm R} (r)\rangle$  and transverse force $F_{\rm T} (r, \phi)$ can be defined as

\begin{equation}
\langle F_{\rm R} (r)\rangle \equiv \frac{{\rm d}\Phi_0(r)}{{\rm d}r} ,\;\;{\rm and}\;\; F_{\rm T} (r, \phi)\equiv \left \lvert\frac{1}{r}\frac{\partial \Phi(r,\phi)}{\partial \phi}\right \rvert
\end{equation}

\noindent where $\Phi_0$ is the $m = 0$ Fourier component of the gravitational potential \citep{Buta2001}. The maximum transverse force to the mean radial force is defined as

\begin{equation}
Q_{\rm T}(r) \equiv \frac{F_{\rm T}^{\rm max}(r)}{\langle F_{\rm R} (r)\rangle}
\end{equation}

\noindent where the maximum tangential force $F_{\rm T}^{\rm max}(r)$ is the maximum of $F_{\rm T} (r, \phi)$ along $\phi$. The ratio map $Q_{\rm T}(r,\phi)$ of a barred galaxy typically presents four thick slabs and the four peaks along these slabs are observed near the four corners of the bar in the deprojected image of the galaxy (see Fig. 1 in \citealt{Lee2020}). The ratio map can be expressed in Cartesian coordinates $Q_{\rm T}(x,y)$ too, where a butterfly-shaped pattern is the typical signature of the presence of a bar. 

The radial profile of $Q_{\rm T}$ is calculated to be averaged over the azimuthal angle $\phi$ \citep{Buta2001}. The analysis of the shape of the $\langle Q_{\rm T}\rangle$ radial profile allows to constrain the bar radius. The location of a peak or a plateau in the $\langle Q_{\rm T}\rangle$ radial profile is adopted as a solid estimate of the bar radius, \rqb. At this radius, four peaks appear in the $Q_{\rm T}(R_{\rm Qb})$ azimuthal profile confirming the correct identification of the bar. 

The method was originally proposed to perform a morphological classification, because the specific characteristics of both the $\langle Q_{\rm T}\rangle$ radial and $Q_{\rm T}(r_{\rm Qb})$ azimuthal profiles allow to distinguish between barred and unbarred galaxies. In particular, \citet{Lee2020} identify a barred galaxy when the $Q_{\rm T}(R_{\rm Qb})$ azimuthal profile presents four peaks corresponding to the four wings of the butterfly-shaped pattern shown in \citet{Buta2001}, together with a global bar strength $Q_{\rm b}>0.15$, defined as the bar force ratio in the polar coordinates

\begin{equation}
    Q_{\rm b} \equiv \frac{1}{n}\sum^n_{i=1} Q_{{\rm T},i}
\end{equation}

\noindent where $Q_{{\rm T},i}$ is the maximum value at each peak on the $Q_{\rm T}(R_{\rm Qb})$ azimuthal profile, and $n$ is the number of the peaks which is equal to four for a bar.

First of all, we considered the SDSS $i$-band images after measuring and subtracting the residual sky level, as done in \citet{morelli2016}. We deprojected the SDSS $i$-band images using the disc parameters provided by \cite{Aguerri2015} and \cite{Cuomo2019b} and reported in Table~\ref{tab:properties}. To double check if these disc parameters were carefully identified and are suitable for the deprojection of the images, we repeated the image deprojection using the disc parameters obtained at the half of the radius of the isophote at a surface brightness level of $\mu_{\rm B}=25$~mag arcsec$^{-2}$ ($R_{25}/2$, RC3). This corresponds to the maximum extension of the residual sky-subtracted SDSS images. The two deprojected images provide consistent results in the resulting analysis. We adopted and present in the following the results corresponding to the deprojection based on the data from \citet{Aguerri2015} and \citet{Cuomo2019b}.

\begin{figure*}[!h]
    \centering
    \includegraphics[scale=0.8]{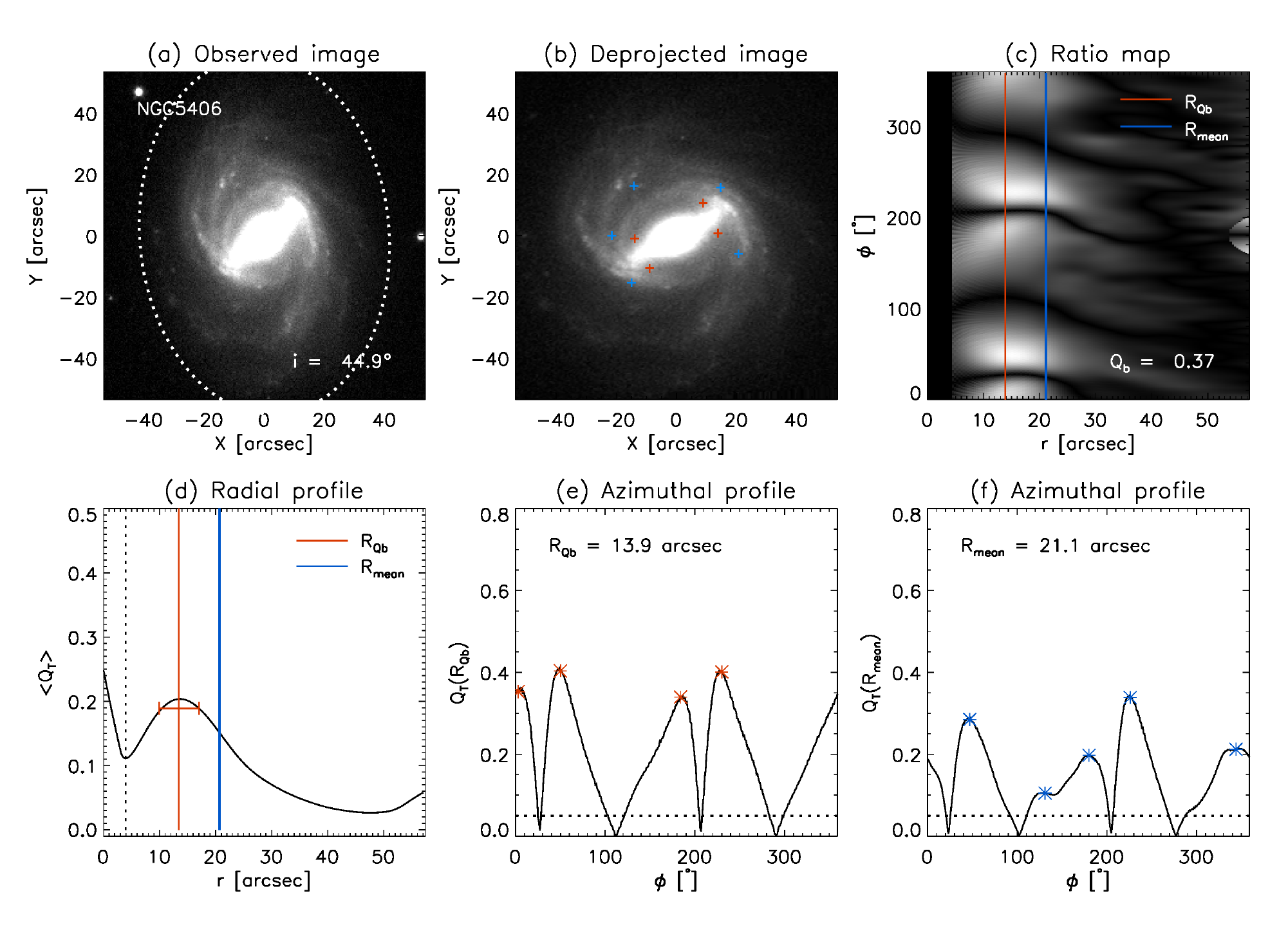}
    \caption{Analysis of the ratio map of NGC~5406. {\em Panel (a):\/} Observed $i$-band SDSS image of the galaxy with north up and east left. The dotted ellipse marks the region of the galaxy considered for the deprojection. The galaxy inclination is given. {\em Panel (b):\/} Deprojected image of the galaxy. The red and blue crosses correspond to the peaks measured in the $Q_T(R_{\rm Qb})$ and $Q_T(R_{\rm mean})$ azimuthal profiles, respectively. {\em Panel (c):\/} Map of the transverse-to-radial force ratio $Q_T(r, \phi)$. The vertical blue and red lines mark the location of \rqb\ and \rmean, respectively. The bar strength is given. {\em Panel (d):\/} Radial profile of $\langle Q_T \rangle$. The vertical dotted line corresponds to the boundary of the bulge-dominated region, identified as the range from the centre to the minimum (or a change in the slope) in the $\langle Q_T \rangle$ radial profile. The vertical blue and red lines mark the location of \rqb\ and \rmean, respectively. The horizontal red segment is the error associated to \rqb. {\em Panel (e):\/} Azimuthal profile of $Q_T(R_{\rm Qb})$. The local maxima of the profile are highlighted by red asterisks. The horizontal dotted line corresponds to the threshold value at $Q_{\rm T} = 0.05$, adopted to count the number of peaks associated to the bar. The value of \rqb\ from this paper is given. {\em Panel (f):\/} Azimuthal profile of $Q_T(R_{\rm mean})$. The local maxima of the profile are highlighted by blue asterisks. The horizontal dotted line corresponds to the threshold value at $Q_{\rm T} = 0.05$, adopted to count the number of peaks associated to the bar. The value of \rmean\ from literature is given.}
\label{fig:pt_mapN5406}
\end{figure*}

We recovered an independent measurement of bar radius \rqb\ and bar strength $Q_{\rm b}$, which are tabulated in Table~\ref{tab:potential_map}. The corresponding error on \rqb\ was obtained as the width of the peak in the $\langle Q_{\rm T}\rangle$ radial profile, calculated where $\langle Q_{\rm T}\rangle$ reaches 95\% of the peak value and using only the right side of the profile, since the peak is not always well defined. The results obtained for our sample galaxies are showed in Figs.~\ref{fig:pt_mapN5406} and \ref{fig:pt_map1} and described in Sect.~\ref{app1}. In particular, Fig.~\ref{fig:pt_mapN5406} shows the analysis of the ratio map for NGC~5406. The original SDSS image, the deprojected one obtained from the disc parameters tabulated in Table~\ref{tab:properties}, and the ratio map as a function of the polar coordinates $(r,\phi)$  are presented in the upper panels. The four thick slabs corresponding to the bar are visible in the inner part of the map (clearly extending up to $r=20$ arcsec). Outermost the slabs transform into a more complex pattern, corresponding to the spiral arms. In the lower panels of the figure are presented the $\langle Q_{\rm T}\rangle$ radial, $Q_{\rm T}(R_{\rm Qb})$ azimuthal, and $Q_{\rm T}(R_{\rm mean})$ azimuthal profiles. 

The galaxies IC~1528 and NGC~5971 are very inclined so the $Q_{\rm T}(r,\phi)$ map analysis is not conclusive. In particular, IC~1528 does not present the typical features of a barred galaxy after deprojection, while the bar of NGC~5971 appears as an artifact structure elongated along the disc minor axis due to deprojection. We decided to discard these galaxies from further analysis of the ratio map and our final sample reduces to 10 objects.

\begin{table}
\caption[Properties of the galaxies.]{Bar radius \rqb\ and strength $Q_{\rm b}$ from the analysis of the transverse-to-radial force ratio map of the sample galaxies.}
    \centering
    \begin{tabular}{ccccc}
    \hline\hline
Galaxy & \rqb & $Q_{\rm b}$ & Class & \rr$_{\rm new}$ \\
 & [arcsec] & & \\
 (1) & (2) & (3) & (4) & (5)\\ 
\hline
IC~1683 & $13.86\pm3.56$ & 0.28 & B & 1.42$^{+0.62}_{-0.83}$\\
IC~5309 & $7.13\pm3.56$ & 0.11 & NB & 0.65$^{+0.47}_{-0.83}$\\
NGC~36 & $12.28\pm2.77$ & 0.33 & B & 1.03$^{+0.37}_{-0.41}$\\
NGC~2553 & $12.68\pm2.77$ & 0.25 & B & 0.90$^{+0.18}_{-0.22}$\\
NGC~2880 & $9.90\pm3.17$ & 0.19 & B & 0.95$^{+0.27}_{-0.39}$\\
NGC~5205 & $12.28\pm3.56$ & 0.29 & B & 0.92$^{+0.22}_{-0.28}$\\
NGC~5406 & $13.86\pm3.56$ & 0.37 & B & 0.79$^{+0.29}_{-0.24}$\\
NGC~5947 & $8.32\pm1.98$ &  0.26 & B & 0.70$^{+0.26}_{-0.30}$\\
NGC~6497 & $11.49\pm2.38$ & 0.34 & B & 0.48$^{+0.16}_{-0.17}$\\
UGC~3253 & $11.88\pm1.89$ & 0.40 & B & 1.00$^{+0.27}_{-0.24}$\\
\hline
    \end{tabular}
    \\
\tablefoot{(1) Galaxy name. (2) Deprojected bar radius. (3) Bar strength. (4) Barred (B) or unbarred (NB) classification according to \cite{Lee2020} criteria. (5) Bar rotation rate estimated as the ratio between \rcor\ tabulated in Table~\ref{tab:properties} and \rqb\ from col. (2).}
    \label{tab:potential_map}
\end{table}

As a first result, we confirmed that all these galaxies host a strong bar, according to the criteria proposed by \citet[][]{Lee2020}, except for IC~5309. This galaxy presents the typical four peaks in the $Q_{\rm T}(R_{\rm Qb})$ azimuthal profile but a lower value of $Q_{\rm b}$ than the threshold adopted to define a barred galaxy. This allowed us to confirm that it is a weakly barred galaxy, as already pointed out by the visual classification from CALIFA \citep{Walcher2014} and the analysis from \citet{Cuomo2019b}. 

Moreover, we tested the value of \rmean\ tabulated in Table~\ref{tab:properties} to recover \rr\ by analysing the $Q_{\rm T}(R_{\rm mean})$ azimuthal profile to look for the typical four peaks expected for the bar. We checked where the resulting peaks are located on the galaxy image (see {\em Panel (b)\/} of Figs.~\ref{fig:pt_mapN5406} and \ref{fig:pt_map1}). 

We found that \rmean$>$\rqb\ by $\sim45$\% on average and in three galaxies the two values are not even consistent to each other within their errors. In fact, \rmean\ is always larger than \rqb\ at face values, but the large errors associated to \rmean\ make the two values consistent in many cases. Only for IC~1683, IC~5309, and NGC~2880, \rmean\ corresponds to the bar radius, but it remains larger than \rqb\ at face values. On the contrary, we realized that \rmean\ for NGC~2553 and NGC~5406 is actually the radius of the ring circling the bar. The corresponding $Q_{\rm T}(R_{\rm mean})$ azimuthal profile shows more than the four peaks we expect to have for a bar. For NGC~36, NGC~5205, NGC~5947, NGC~6497, and UGC~3253, the $Q_{\rm T}(R_{\rm mean})$ azimuthal profile shows four peaks but the value of \rmean\ corresponds to a galactocentric distance where a ring or spiral arms are clearly visible in the galaxy image. A detailed description of the morphology of the sample galaxies with a comparison between \rmean\ and \rqb\ is presented in Sect.~\ref{app1}. 

We calculated \rr\ by dividing the value of \rcor\ tabulated in Table~\ref{tab:properties} by \rqb\ and present the results in Table~\ref{tab:potential_map}. The value of \rr\ increases since \rqb\ is shorter than \rmean. All the galaxies move to the fast regime, except for NGC~6497. We discuss this galaxy later and conclude it does not host an ultrafast bar. In addition, the bars of IC~1683 and NGC~36 are consistent with being slow.

\subsection{Comparison of bar radius estimates}

\begin{figure*}[!h]
    \centering
    \includegraphics[scale=0.83]{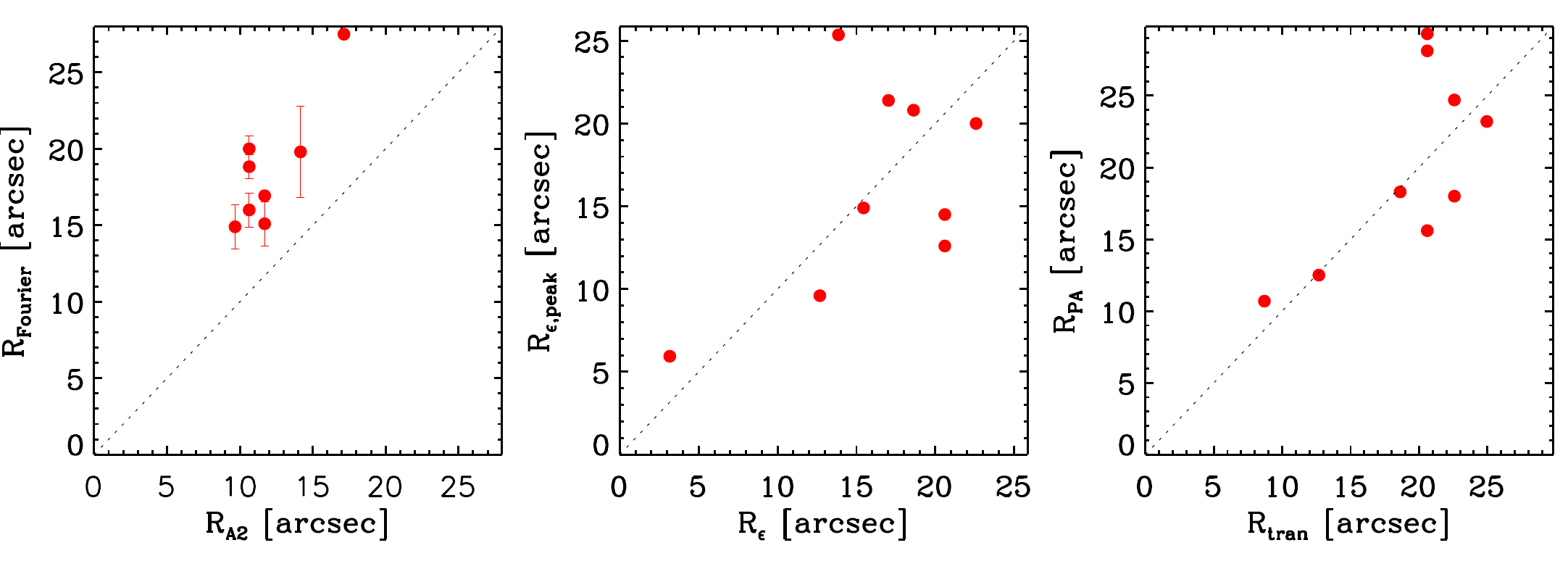}
    \caption{Comparison between the bar radius of the sample galaxies obtained with similar methods based on Fourier analysis (left panel), ellipticity (central panel) and PA (right panel) radial profiles.}
    \label{fig:bar_lengths_comp1}
\end{figure*}

We measured \rbar\ in the sample galaxies with three more methods based on the ellipse fitting and Fourier analysis of the deprojected images, as done by \citet{Lee2019,Lee2020}. In particular, we calculated $R_\epsilon$ and $R_{\rm tran}$ from the ellipticity and PA radial profiles of the deprojected SDSS $i$-band images, as the radius where the maximum of $\epsilon$ occurs and where the PA varies by 2\degr\ with respect to the location of the peak in $\epsilon$, respectively and $R_{\rm A_2}$ as the radius corresponding to the maximum value of the amplitude $A_2$ of the Fourier $m=2$ component, in the region where the phase angle $\phi_2$ remains constant. We compare the new \rbar\ estimates with those obtained with similar methods by \cite{Aguerri2015} and \cite{Cuomo2019} in Fig.~\ref{fig:bar_lengths_comp1}. In particular, we put $R_\epsilon$ together with $R_{\rm \epsilon,peak}$, $R_{\rm tran}$ with $R_{\rm PA}$, and $R_{\rm A_2}$ with $R_{\rm Fourier}$. While $R_{\rm \epsilon,peak} \sim R_\epsilon$ and $R_{\rm PA} \sim R_{\rm tran}$ although with some scatter, it results that systematically $R_{\rm Fourier}>R_{\rm A_2}$. This discrepancy is due to the slightly different definition of \rbar\ in the two Fourier-based methods. In fact, $R_{\rm A_2}$ considers the peak of the $m=2$ component of the Fourier series in the region where $\phi_2$ remains constant to exclude the range with higher peaks in the Fourier $m=2$ component caused by spiral arms. On the other, hand, $R_{\rm Fourier}$ requires the higher orders of the Fourier series and do not include the behaviour of the corresponding phase angles. The even components of the Fourier series, together with the corresponding phase angles and in particular the $m=2$ one, can be strongly affected by the prominence of the bulge \citep{Debattista2002,Lee2020}.

\begin{figure*}[!h]
    \centering
    \includegraphics[scale=0.95]{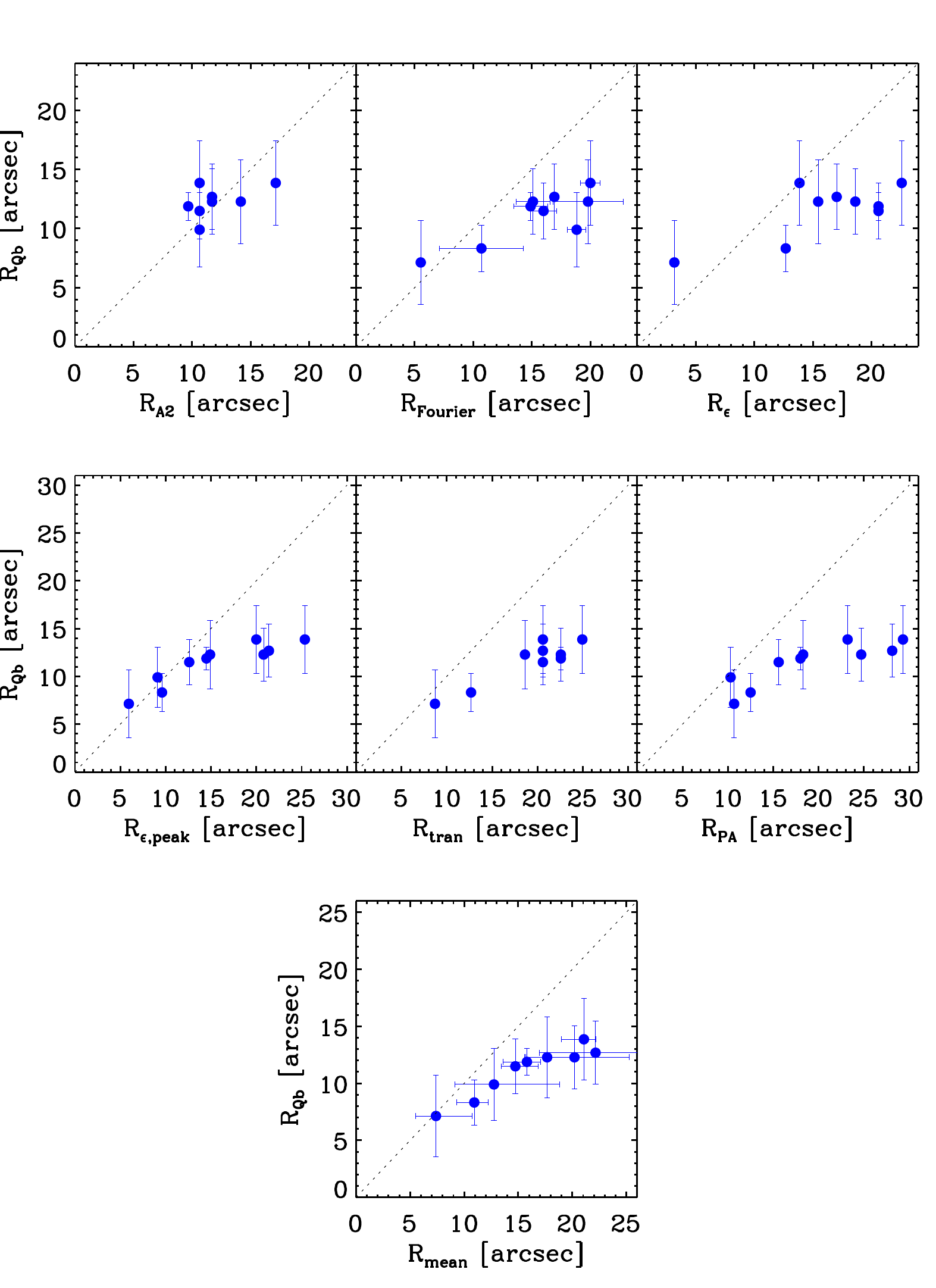}
    \caption{Comparison between the bar radius of the sample galaxies obtained with several methods and the analysis of the ratio maps. The bottom panel shows the comparison between \rmean\ and \rqb\ given in Tables~\ref{tab:properties} and \ref{tab:potential_map}, respectively.}
    \label{fig:bar_lengths_comp2}
\end{figure*}

In Fig.~\ref{fig:bar_lengths_comp2} we compare all the different \rbar\ estimates of the sample galaxies to \rqb\ as well as their mean value \rmean\ given in Table~\ref{tab:properties}. We notice that \rqb\ is systematically shorter than \rmean\ and the other available measurements of \rbar. In particular, \rqb\ is always shorter than \rmean\ at face values, but for three galaxies the two \rbar\ estimates are consistent within the errors. This is due to the large errors associated to \rmean. The difference between \rmean\ and \rqb\ increases for the longer bars of the sample, like those residing in IC~1683, NGC~36, NGC~2553, and NGC~5406. Our findings are in agreement with those of \citet{Lee2020} who compared their measurements of \rbar\ in sample of about 400 spiral galaxies with those available in literature \citep{Laurikainen2002, DiazGarcia2016} and found a strong correlation between $R_{\rm A_2}$ and \rqb. 

As a consequence of the trends shown in Fig.~\ref{fig:bar_lengths_comp2}, we can conclude that the various estimates of \rbar\ generally lead to smaller values of \rr\ with respect to \rqb, although a solid estimate of the errors on \rbar\ is not available so far for all the measurement methods. The values of \rr\ derived for \rmean\ and \rqb\ are given in Table~\ref{tab:properties} and \ref{tab:potential_map}, respectively, while the values of \rr\ from $R_{\rm \epsilon,peak}$, $R_{\rm PA}$, $R_{\rm Fourier}$, $R_\epsilon$, $R_{\rm tran}$, and $R_{\rm A_2}$ are reported in Table~\ref{tab:rotation_rates}.

\subsection{Results for individual galaxies}
\label{app1}

We performed the morphological classification of the sample galaxies following the criteria given by \citet{Buta2015}, \citet{Buta2017}, and \citet{Bittner2020}. \\

\noindent \textbf{IC~1528} is a flocculent galaxy with strong spiral arms and no rings. Due to its high inclination ($i\sim70\degr$), we can not understand whether it is an unbarred galaxy, where the bar is mimicked by the winding spiral arms, or a genuine weakly barred galaxy, as suggested by the weak four peaks in the $Q_{\rm T}(R_{\rm Qb})$ azimuthal profile. The ratio map does not present the typical four slabs for barred galaxies and there is no peak or plateau in the $\langle Q_{\rm T}\rangle$ radial profile. Therefore, we excluded IC~1528 from the discussion related to the analysis of the ratio maps.\\

\noindent \textbf{IC~1683} is a two-armed spiral galaxy without rings, which we classified to belong to the (s) variety with arms breaking out directly from the bar. The ratio map presents four slabs for $r\lesssim35$ arcsec and an outer complex pattern corresponding to the spiral arms. The $\langle Q_{\rm T}\rangle$ radial profile decreases till a local minimum at $r=4.0$ arcsec in the bulge region, then it increases to a maximum value at \rqb. The corresponding $Q_{\rm T}(R_{\rm Qb})$ azimuthal profile presents four peaks with $Q_{\rm b}=0.28$. The $Q_{\rm T}(R_{\rm mean})$ azimuthal profile shows four weaker peaks. The value of \rmean\ is larger than \rqb\ and they are not consistent with each other.\\

\noindent \textbf{IC~5309} is an inclined spiral galaxy with strong grand-design spiral arms but no rings. The ratio map presents four slabs for $r\lesssim10$ arcsec and an outer complex pattern corresponding to the spiral arms. The $\langle Q_{\rm T}\rangle$ radial profile is dominated by the bulge till $r=2.8$ arcsec and then decreases to a kind of a plateau, corresponding to the four peaks in the $Q_{\rm T}(R_{\rm Qb})$ azimuthal profile. We classified the galaxy as weakly barred for its low bar strength $Q_b=0.11$. The $Q_{\rm T}(R_{\rm mean})$ azimuthal profile presents only three peaks. The values of \rmean\ and \rqb\ are consistent within the errors.\\

\noindent \textbf{NGC~36} is a grand-design spiral galaxy with a inner ring and an outer pseudo-ring which we highlighted by adopting the (r) and (R$'_2$) varieties for the morphological classification. The ratio map presents four slabs extending to the outer regions of the galaxy and corresponding to both the bar and inner ring. The $\langle Q_{\rm T}\rangle$ radial profile decreases to a local minimum at $r=4.0$ arcsec in the bulge region, then it increases to a maximum value at \rqb. The corresponding $Q_{\rm T}(R_{\rm Qb})$ azimuthal profile presents four peaks with $Q_{\rm b}=0.33$. The $Q_{\rm T}(R_{\rm mean})$ azimuthal profile shows four peaks, even if they are less prominent. The size of the inner ring corresponds to \rmean, which is not consistent with \rqb.\\

\noindent \textbf{NGC~2553} is a two-armed spiral which we recognize to have an inner (r) and an outer (R$_1$R$'_2$) morphology. The ratio
maps present four well-defined slabs extending to the outer regions of the galaxy. The $\langle Q_{\rm T}\rangle$ radial profile decreases to a local minimum at $r=3.6$ arcsec in the bulge region, then it increases till a maximum value at \rqb. The corresponding $Q_{\rm T}(R_{\rm Qb})$ azimuthal profile presents four peaks with $Q_{\rm b}=0.26$. The $Q_{\rm T}(R_{\rm mean})$ azimuthal profile shows six weak peaks. The size of the inner ring corresponds to \rmean, which is consistent with \rqb\ within the
errors.\\

\noindent \textbf{NGC~2880} is the only lenticular galaxy of the sample. It hosts a large bulge and a bar almost aligned with the disc minor axis. The ratio map presents four well-defined slabs extending to the outer regions of the galaxy. The $\langle Q_{\rm T}\rangle$ radial profile decreases to a local minimum at $r=6.0$ arcsec in the bulge region. Then, it increases to a plateau corresponding to the four peaks in the $Q_{\rm T}(R_{\rm Qb})$ azimuthal profile with $Q_{\rm b}=0.19$. The $Q_{\rm T}(R_{\rm mean})$
azimuthal profile shows the same four peaks. The values of \rmean\ and \rqb\ are consistent within the errors.\\

\noindent \textbf{NGC~5205} is a multiple-armed spiral galaxy with an inner broken ring which we translated into an (rs) classification. The ratio map presents four slabs for $r\lesssim30$ arcsec and an outer complex pattern corresponding to the spiral arms. The $\langle Q_{\rm T}\rangle$ radial profile decreases to a local minimum at $r=5.9$ arcsec in the bulge region, then it increases to a maximum value at \rqb. The corresponding $Q_{\rm T}(R_{\rm Qb})$ azimuthal profile presents four peaks with $Q_{\rm b}=0.29$. The peaks are nearly the same in the $Q_{\rm T}(R_{\rm mean})$ azimuthal profile. The size of the inner broken ring corresponds to \rmean, which is consistent with \rqb\ within the errors.\\

\noindent \textbf{NGC~5406} is a multiple-armed spiral galaxy with an inner broken ring and an (rs) morphology. The ratio map presents four slabs for $r\lesssim25$ arcsec and an outer complex pattern corresponding to the spiral arms. The $\langle Q_{\rm T}\rangle$ radial profile decreases to a local minimum at $r=4.4$ arcsec in the bulge region, then it increases till a maximum value at \rqb. The corresponding $Q_{\rm T}(R_{\rm Qb})$ azimuthal profile presents four peaks with $Q_{\rm b}=0.37$. The value of \rmean\ corresponds to the ring size and the presence of a spiral arm in the ring region gives rise to the fifth peak observed in the $Q_{\rm T}(R_{\rm Qb})$ azimuthal profile. The values of \rmean\ and \rqb\ are consistent within the errors.\\

\noindent \textbf{NGC~5947} is a multiple-armed spiral galaxy. We noticed the presence of an inner ring, which translates into an (r) classification. The ratio map presents four slabs for $r\lesssim20$ arcsec and an outer complex pattern corresponding to the spiral arms. The $\langle Q_{\rm T}\rangle$ radial profile decreases to a local minimum at $r=2.8$ arcsec in the bulge region, then it increases to a maximum value at \rqb. The corresponding $Q_{\rm T}(R_{\rm Qb})$ azimuthal profile presents four peaks with $Q_{\rm b}=0.26$. The $Q_{\rm T}(R_{\rm mean})$ azimuthal profile is characterised by four peaks too. The size of the ring corresponds to \rmean, which is consistent with \rqb\ within the errors.\\

\noindent \textbf{NGC~5971} is an highly-inclined spiral galaxy ($i\sim70\degr$) with multiple arms. Its deprojection produces an artifact bar structure along the disc minor axis hampering the analysis of the ratio map, which nevertheless shows the typical four slabs associated to a bar. We excluded NGC~5971 from the discussion related to the analysis of the ratio maps.\\

\noindent \textbf{NGC~6497} is a multiple-armed spiral galaxy with flocculent spiral arms. We translated the presence of outer rings into an (R$_1$R$'_2$) morphology. The ratio map presents four slabs for $r\lesssim15$ arcsec. The $\langle Q_{\rm T}\rangle$ radial profile decreases to a local minimum at $r=3.6$ arcsec in the bulge region, then it increases to a maximum value at \rqb. The corresponding $Q_{\rm T}(R_{\rm Qb})$ azimuthal profile presents four peaks with $Q_{\rm b}=0.34$. The same peaks are shown by the $Q_{\rm T}(R_{\rm mean})$ azimuthal profile. The ring size is consistent with \rmean, which agrees with \rqb\ within the errors.\\

\noindent \textbf{UGC~3253} is a multiple-armed spiral galaxy. The presence of an inner ring translates into an (r) classification. The ratio map presents four slabs for $r\lesssim20$ arcsec and an outer complex pattern corresponding to the spiral arms.  The $\langle Q_{\rm T}\rangle$ radial profile decreases to a local minimum at $r=4.0$ arcsec in the bulge region, then it increases to a maximum value at \rqb. The corresponding $Q_{\rm T}(R_{\rm Qb})$ azimuthal profile presents four peaks with $Q_{\rm b}=0.40$. The $Q_{\rm T}(R_{\rm mean})$ azimuthal profile shows the same peaks. The size of the inner ring corresponds to \rmean, which is consistent with \rqb\ within the errors.\\

\section{Discussion}
\label{sec:discussion}

The bars of a number of disc galaxies of the CALIFA survey with direct and accurate measurements of \omegabar\ are characterised by ${\cal R}<1$ \citep{Aguerri2015, Cuomo2019b}. These unexpected observational findings are dynamically incompatible with the theoretical predictions about the stability of stellar orbits supporting the bar \citep{Contopoulos1981}. Therefore, we decided to test whether these ultrafast bars are actually an artefact due to an overestimated value of \rbar\ and/or an underestimated value of \rcor\ rather than a new class of non-axisymmetric stellar components, whose orbital structure has not been yet understood.

For all the sample galaxies, we found that the \rbar\ measurement based on the analysis of the ratio maps is shorter than that obtained with other methods based on ellipse fitting and Fourier analysis of the deprojected galaxy image. These methods turned out to be quite sensible to the presence of rings, pseudo-rings, and spiral arms which are very common in the galaxies we analysed and lead to systematically larger values of \rbar. All the sample galaxies present a rather complex spiral morphology, except for NGC~2880. This is a lenticular galaxy, which does not show any additional component to the bulge, bar, and disc and its \rqb\ is consistent (although smaller) with \rbar\ derived from other methods.

When adopting \rqb\ to calculate \rr, all the galaxies turned out to host a fast bar at 95\% confidence level with the only exception represented by NGC~6497. This galaxy was previously discussed in detail by \citet{Aguerri2015} and \citet{Garma-Oehmichen2020}. \citet{Aguerri2015} analysed the galaxy extinction map to rule out problems in measuring \rbar\ due to dust and considered the gas kinematics to check \vcirc\ and hence \rcor\ obtained from stellar dynamics. \citet{Garma-Oehmichen2020} reassessed the error budget of \omegabar\ by considering a broader set of error sources affecting the TW method. They remeasured \omegabar\ and recalculated \rcor\ obtaining ${\cal R}=1.08^{+0.31}_{-0.25}$ which makes NGC~6497 fully consistent with the fast-bar regime.

Our results on the problems in measuring \rbar\ in barred galaxies confirm previous findings based on the analysis of mock images and numerical simulations. In fact, \cite{MichelDansac2006} showed that \rr\ can increase from 1.0 to 1.4 just by changing of the method adopted to recover \rbar. Using simulated images, they showed that \rbar\ obtained from the location of the maximum in the $\epsilon$ radial profile is closer to \rcor, whereas \rbar\ estimated from the location of the constant PA or from the Fourier analysis correlates with the ultra-harmonic 4:1 resonance well within \rcor. Different \rbar\ values translates into different \rr\ estimates. More recently, \cite{Petersen2019} have shown with $N$-body simulations that \rbar\ measured from ellipse fitting overestimates by a factor 1.5--2 the radial extent of the bar recovered from the maximum excursion of $x_1$ stellar orbits. This is because many untrapped stellar orbits reside in the physical regions of the $x_1$ family and are considered part of the bar by the ellipse fitting. 

Using the images of mock galaxies, \cite{Lee2020} showed that the measurement of \rbar\ based on the ratio map are overestimated when the bulge-to-total ratio $B/T$ increases from 0 to 0.8, but the same effect was pointed out for the ellipse fitting and Fourier analysis methods as well. This means that if there is a substantial contribution of the bulge, the corresponding rotation rate is more effectively driven out the ultrafast regime into the fast one. The bulges in our galaxies give a relatively low contribution to the total luminosity, with a mean value of $\langle B/T\rangle=0.15$ \citep{MendezAbreu2017}, when excluding the SB0 NGC~2880 which hosts a large bulge ($B/T=0.46$). Moreover, the ratio map allows to disentangle the radius corresponding to the maximum strength of the bar from that of the spiral arms and/or of rings by comparing the $Q_{\rm T}(R_{\rm Qb})$ and $Q_{\rm T}(R_{\rm mean})$ azimuthal profiles and looking where the local maxima of the azimuthal profiles are located in the image of the galaxy. As a future perspective, it is worth checking whether bars located in the safe fast regime are in turn affected by a wrong estimate of bar radius, especially when the host galaxies present spiral arms and/or rings. If this is the case, some of those bars may indeed belong to the slow regime, challenging the conclusions drawn so far in the framework of bar rotation regimes. Unfortunately, all the most widely adopted methods to recover \rbar\ are based only on photometry without considering kinematic information, which may help to successfully constrain the extension of the bar.

\cite{Hilmi2020} used hydro-dynamical simulations of Milky Way-like galaxies to assess the variation of the bar parameters on a dynamical timescale due to the interaction with the spiral structure. Using different approaches based on photometry, they recovered \rbar, \sbar, and \omegabar\ and traced their evolution with time. All the adopted methods overestimate \rbar. The bar rotates faster than the spiral pattern and sometimes bar and spiral arms overlap. When the bar is connected to spiral arms, it seems to increase its radius. These bar pulsations are due to the coupling with the modes of the spiral pattern. Since the spiral modes can be odd, the two bar ends typically do not connect at exactly the same time to a spiral arm, so the two bar {\rm radii} (one per each half) may be different at some given time. According to \cite{Hilmi2020}, in $\sim 50\%$ of Milky Way-like galaxies, the \rbar\ measurements of SBab and SBbc galaxies are overestimated by $\sim 15\%$ and $\sim 55\%$, respectively with the stronger bars driving larger errors. Considering the sample analysed by \cite{Cuomo2020}, we point out that ultrafast bars seem to be associated to stronger bars when only galaxies with TW measurements are considered. We found that \rqb\ values are on average $\sim45\%$ shorter than other \rbar\ estimates for our sample galaxies and that this difference decreases from SBab to SBc galaxies. Moreover, \cite{Hilmi2020} showed that while the bar is increasing its radius due to the interaction with the spiral structure, the corresponding \omegabar\ decreases, but at a lower rate. Therefore, the two effects do not cancel out: intrinsically fast bars may appear as ultrafast. Given that \omegabar\ is well determined in our sample galaxies with the TW method, finding ultrafast bars need to be associated to an erroneous determination of \rbar.

It should be noticed that it is mandatory to adopt the same approach in measuring \rbar\ when the theoretical predictions \citep{weinberg1985,Hernquist1992} and numerical simulations \citep{Debattista2000,Zou2019,Ghafourian2020} are compared to observational results to avoid misinterpreting the data. On the other hand, \rr\ also depends on \rcor\ and hence on \omegabar\ and \vcirc. \citet{Aguerri2015} and \citet{Cuomo2019b} derived \vcirc\ from the asymmetric drift-corrected stellar kinematics in the disc region \citep{Binney2008} and verified that their values agree with the Tully-Fisher relation predictions \citep{Tully1977,Reyes2011}. Moreover, \citet{Aguerri2015} recovered \vcirc\ for NGC~36, NGC~5205, and NGC~6497 using available gas kinematics \citep{Theureau1998,Garcia-Lorenzo2015} and excluded in these cases the determination of \vcirc\ can explain the observed ultrafast regime. On the other hand, \citet{Garma-Oehmichen2020} directly estimated the value of \rcor\ as the intersection between \omegabar\ and the modelled angular rotation curve, which is useful for galaxies where the rotation curve rises slowly and \rcor\ can be overestimated when measured using \vcirc, but they did not infer any conclusion about the ultrafast regime. 

In this paper, we considered only ultrafast barred galaxies with a direct measurement of \omegabar\ from the TW method. We already discussed the sources of error of the TW method in Sect.~\ref{sec:results}, but some further considerations are worth to be done.

Using $N$-body simulations, \cite{Zou2019} suggested that ${\cal R}<1$ can occur when $\Delta$PA between bar and disc major axes is overestimated, the bar is too close to the disc minor axis, and the field of view is too small to guarantee the convergence of the TW integrals. In addition, \cite{Cuomo2019} showed that ${\cal R}<1$ could be also the result of a wrong estimate of the disc PA from ellipse fitting when the PA radial profile is not constant as for warped discs. All the findings imply that ultrafast bars could be due to a wrong application of the TW method. However, we exclude this is the case for our sample galaxies which were carefully selected to be perfectly suitable for the correct application of the TW technique. As detailed by \citet{Aguerri2015} and \citet{Cuomo2019b} in performing their TW measurements, the PA of the bar and disc major axes were carefully measured with an ellipse fitting analysis, the constant portion of the PA profile corresponding to the disc region where the bar is located was identified, no correlation was found between the relative errors of \rcor\ and \rbar\ and the values of the disc $i$ and bar $\Delta$PA with respect to the disc major axis, and the radial extent of the photometric and kinematic data was chosen to allow the convergence of the TW integrals.

The existence of distinct pattern speeds corresponding to different galaxy structures was extensively discussed \citep{Rautiainen2008,Cuomo2019b}. There were some efforts to modify and apply the TW method to recover multiple pattern speeds. In fact, the assumption of a well-defined rigidly rotating pattern speed in a barred galaxy can be questioned in the case of rings and/or spiral arms. The bar and spiral arms possibly share the same pattern speed when the arms are driven by the bar \citep{Sanders1976}, or the bar and spiral arms can have different pattern speeds even if they are connected \citep{Sellwood1988,Beckman2018,Hilmi2020}. Moreover, these pattern speeds may vary in space and/or time \citep{Toomre1981,Bertin1996}. The TW method provides an average pattern speed, if multiple pattern speeds are present.
Since the length of the pseudo-slits must reach the axisymmetric disc, crossing both the bar and the spiral arms is unavoidable. As already pointed out by \cite{Tremaine1984}, \cite{Debattista2002b} showed that small perturbations in the disc density do not contribute significantly to \omegabar. Low amplitudes correspond to a rapidly growing structure, which corresponds to spiral arms. \cite{Meidt2008} adapted the TW method to measure different pattern speeds from independent radial regions. They argued that also for spiral galaxies, the bar contribution to the measured pattern speed is maximal when only the photometric and kinematic integrals taken across the bar are considered in the analysis. This is commonly done in the application of the TW method to spiral galaxies \citep{Aguerri2015,Cuomo2019b,Guo2019,Garma-Oehmichen2020,Williams2021}. In this case, the measured pattern speed is reliable \citep{Meidt2008}.
Moreover, deviations from the bar pattern speed are small when the spiral arms are dim \citep{Williams2021}. A slope change of the straight-line fitting the TW integrals was also interpreted as the signature of a nuclear bar rotating with a different pattern speed with respect to the main bar \citep{Corsini2003, Maciejewski2006,Meidt2009}. A slight slope change is observed in some of the galaxies of our sample, but also in other galaxies not hosting an ultrafast bar \citep{Aguerri2015, Cuomo2019b}, so we can conclude there is no clear link between the originally observed ultrafast regime and the slope change.

Dark gaps are commonly seen in early to intermediate-type barred galaxies having inner and outer rings or related features: the radial zone between an inner and outer ring appears dark, either continuously or in 2\,--\,4 distinct sections \citep{Kim2016,James2016,Buta2017}. \cite{Buta2017} suggested that the dark gaps between inner and outer rings are associated with the $L_4$ and $L_5$ Lagrangian points in the gravitational potential of a bar or an oval. In turn, these points are theoretically expected to lie very close to the corotation resonance of the bar pattern, so the gaps may provide the location of \rcor. According to \citet{Kim2016}, the inner disc stars are swept by the bar and thus sparse regions are thought to be produced by the bar driven secular evolution. Pronounced light deficits are expected to be observed as the bar evolves becoming more extended and stronger. Indeed, during the evolution of a barred galaxy, the bar loses angular momentum by trapping nearby disc stars onto elongated orbits. This results into an increase of the bar radius and strength \citep{Athanassoula2002,Athanassoula2013,Kim2016}. \cite{Buta2017} found a mean bar rotation rate $\langle {\cal R} \rangle=1.58\pm0.04$ for a sample of 50 galaxies with dark gaps and this means that they host slow-rotating bars. Measuring ${\cal R}<1$ in galaxies with dark gaps reinforce the idea that ultrafast bars are due to an artifact in the determination of the rotation rate. Our analysis moved ultrafast bars in the fast (and even in the slow) regimes, in agreement with the expected results for evolved galaxies with dark gaps.

However, \cite{Buta2017} identified a sub-class of dark gaps, where the interior of an inner ring appears darker than outside. He found a redder color in the dark gaps with respect to the bars and no recent star formation. This is in agreement with the scenario of a bar depleting nearby regions from stars, while evolving. On the other hand, the rotation rates seem to locate these bars in the ultrafast regime. \cite{Buta2017} claimed that if the dark spaces in these galaxies are interpreted in the same way as for the inner/outer ring gap galaxies, then either the existence of ultrafast bars would have to be acknowledged or another mechanism for forming dark gaps that has nothing to do with Lagrangian points would have to be hypothesized. This specific morphology can be recognised in two of our galaxies, NGC~5406 and NGC~5947. Since we excluded that the analysed bars are rotating extremely fast, we stress a different mechanism is need to explain at least this sub-class of galaxies with dark gaps.

\section{Conclusions}
\label{sec:conclusions}

In this paper we considered the case of ultrafast bars, which are observed in more than 10\% of barred galaxies with a direct measurement of the bar pattern speed. These bars end beyond the corotation radius and therefore challenge our understanding of stellar dynamics in barred galaxies. We aimed at investigating whether ultrafast bars are actually an artefact due to an overestimated value of \rbar\ and/or an underestimated value of \rcor\ rather than a new class of non-axisymmetric stellar components, whose orbital structure has not been yet understood.

We took into account the 12 barred galaxies, for which \omegabar\ was carefully measured by applying the TW method to the integral-field spectroscopic data obtained by the CALIFA survey and turned out to host an ultrafast bar according to its ${\cal R}<1$ \citep{Aguerri2015, Cuomo2019b}.

We checked that the galaxies were selected to be suitable for the application of the TW method and confirmed the values obtained for their \rcor. Then, we analysed the issues related to the available \rbar\ measurements by \citet{Aguerri2015} and \citet{Cuomo2019b} based on the ellipse fitting and Fourier analysis of the deprojected SDSS $i$-band images of the sample galaxies. We also derived new estimates of \rbar\ from the analysis of the $\epsilon$ and PA radial profiles and of the Fourier $m=2$ mode following the prescriptions of \citet{Lee2019,Lee2020}.

We realized that nearly all the sample galaxies are spiral galaxies with an inner ring or pseudo-ring circling the bar and/or strong spiral arms, which hamper the \rbar\ measurements based on the ellipse fitting and Fourier analysis of the deprojected galaxy images. According to these methods, the ends of the ultrafast bars overlap the inner ring and/or the spiral arms making the adopted \rbar\ unreliable.

Hence, we performed a further estimate of \rbar\ using the method proposed by \citet{Lee2020} and based on the analysis of the ratio maps, which we successfully applied to 10 galaxies. These values of \rbar\ are systematically smaller than \rmean\ and become smaller (or similar) to the corresponding \rcor. This implies that the corresponding \rr\ are larger than those obtained before. All the galaxies turned out host a fast bar at 95\% confidence level with the only exception represented by NGC~6497. However, \citet{Garma-Oehmichen2020} have recently recalculated \rcor\ for this galaxy and found it is consistent with the fast-bar regime too.

We can confidently conclude that ultrafast bars are no longer observed when a correct measurement of \rbar\ is adopted. However, we still miss a solid estimate of \rbar\ based on both photometric and kinematic data unveiling the extension of the stellar orbits which support the bar and helping the comparison with theoretical prescriptions and numerical simulations. This task requires further investigation.

\begin{acknowledgements}
We thank the anonymous referee for the constructive report that helped us to improve the paper. We are grateful to V. P. Debattista and T. Kim for their valuable comments. We want to thank J. Mendez-Abreu for providing the processed data. VC acknowledges support from the ESO-Government of Chile Joint Comittee programme ORP060/19. VC, CB, and EMC are supported by MIUR grant PRIN 2017 20173ML3WW$\_$001 and Padua University grants DOR1885254/18, DOR1935272/19, and DOR2013080/20. JALA is supported by the Spanish MINECO grant AYA2017-83204-P. 
\end{acknowledgements}

\bibliographystyle{aa}

\begin{thebibliography}{100}
\expandafter\ifx\csname natexlab\endcsname\relax\def\natexlab#1{#1}\fi

\bibitem[{{Abazajian} {et~al.}(2009){Abazajian}, {Adelman-McCarthy},
  {Ag{\"u}eros}, {Allam}, {Allende Prieto}, {An}, {Anderson}, {Anderson},
  {Annis}, {Bahcall}, \& et~al.}]{Abazajian2009}
{Abazajian}, K.~N., {Adelman-McCarthy}, J.~K., {Ag{\"u}eros}, M.~A., {et~al.}
  2009, \apjs, 182, 543

\bibitem[{{Abolfathi} {et~al.}(2018){Abolfathi}, {Aguado}, {Aguilar}, {Allende
  Prieto}, {Almeida}, {Ananna}, {Anders}, {Anderson}, {Andrews}, {Anguiano}, \&
  et~al.}]{Abolfathi2018}
{Abolfathi}, B., {Aguado}, D.~S., {Aguilar}, G., {et~al.} 2018, \apjs, 235, 42

\bibitem[{{Adams} {et~al.}(1989){Adams}, {Ruden}, \& {Shu}}]{Adams1989}
{Adams}, F.~C., {Ruden}, S.~P., \& {Shu}, F.~H. 1989, \apj, 347, 959

\bibitem[{{Aguerri} {et~al.}(2005){Aguerri}, {Elias-Rosa}, {Corsini}, \&
  {Mu{\~n}oz-Tu{\~n}{\'o}n}}]{aguerri2005}
{Aguerri}, J.~A.~L., {Elias-Rosa}, N., {Corsini}, E.~M., \&
  {Mu{\~n}oz-Tu{\~n}{\'o}n}, C. 2005, \aap, 434, 109

\bibitem[{{Aguerri} {et~al.}(2009){Aguerri}, {M{\'e}ndez-Abreu}, \&
  {Corsini}}]{Aguerri2009}
{Aguerri}, J.~A.~L., {M{\'e}ndez-Abreu}, J., \& {Corsini}, E.~M. 2009, \aap,
  495, 491

\bibitem[{{Aguerri} {et~al.}(2015){Aguerri}, {M{\'e}ndez-Abreu},
  {Falc{\'o}n-Barroso}, {Amorin}, {Barrera-Ballesteros}, {Cid Fernandes},
  {Garc{\'{\i}}a-Benito}, {Garc{\'{\i}}a-Lorenzo}, {Gonz{\'a}lez Delgado},
  {Husemann}, {Kalinova}, {Lyubenova}, {Marino}, {M{\'a}rquez}, {Mast},
  {P{\'e}rez}, {S{\'a}nchez}, {van de Ven}, {Walcher}, {Backsmann},
  {Cortijo-Ferrero}, {Bland-Hawthorn}, {del Olmo}, {Iglesias-P{\'a}ramo},
  {P{\'e}rez}, {S{\'a}nchez-Bl{\'a}zquez}, {Wisotzki}, \&
  {Ziegler}}]{Aguerri2015}
{Aguerri}, J.~A.~L., {M{\'e}ndez-Abreu}, J., {Falc{\'o}n-Barroso}, J., {et~al.}
  2015, \aap, 576, A102

\bibitem[{{Aguerri} {et~al.}(2000){Aguerri}, {Mu{\~n}oz-Tu{\~n}{\'o}n},
  {Varela}, \& {Prieto}}]{Aguerri2000}
{Aguerri}, J.~A.~L., {Mu{\~n}oz-Tu{\~n}{\'o}n}, C., {Varela}, A.~M., \&
  {Prieto}, M. 2000, \aap, 361, 841

\bibitem[{{Athanassoula}(1992)}]{Athanassoula1992}
{Athanassoula}, E. 1992, \mnras, 259, 345

\bibitem[{{Athanassoula}(2003)}]{Athanassoula2003}
{Athanassoula}, E. 2003, \mnras, 341, 1179

\bibitem[{{Athanassoula}(2014)}]{Athanassoula2014}
{Athanassoula}, E. 2014, \mnras, 438, L81

\bibitem[{{Athanassoula} {et~al.}(2013){Athanassoula}, {Machado}, \&
  {Rodionov}}]{Athanassoula2013}
{Athanassoula}, E., {Machado}, R. E.~G., \& {Rodionov}, S.~A. 2013, \mnras,
  429, 1949

\bibitem[{{Athanassoula} \& {Misiriotis}(2002)}]{Athanassoula2002}
{Athanassoula}, E. \& {Misiriotis}, A. 2002, \mnras, 330, 35

\bibitem[{{Barazza} {et~al.}(2008){Barazza}, {Jogee}, \&
  {Marinova}}]{barazza2008}
{Barazza}, F.~D., {Jogee}, S., \& {Marinova}, I. 2008, \apj, 675, 1194

\bibitem[{{Beckman} {et~al.}(2018){Beckman}, {Font}, {Borlaff}, \&
  {Garc{\'\i}a-Lorenzo}}]{Beckman2018}
{Beckman}, J.~E., {Font}, J., {Borlaff}, A., \& {Garc{\'\i}a-Lorenzo}, B. 2018,
  \apj, 854, 182

\bibitem[{{Bertin} \& {Lin}(1996)}]{Bertin1996}
{Bertin}, G. \& {Lin}, C.~C. 1996, {Spiral structure in galaxies a density wave
  theory}

\bibitem[{{Binney} \& {Tremaine}(2008)}]{Binney2008}
{Binney}, J. \& {Tremaine}, S. 2008, {Galactic Dynamics: Second Edition}
  (Princeton University Press, Princeton, NJ USA)

\bibitem[{{Bittner} {et~al.}(2020){Bittner}, {Gadotti}, {Elmegreen},
  {Athanassoula}, {Elmegreen}, {Bosma}, \& {Mu{\~n}oz-Mateos}}]{Bittner2020}
{Bittner}, A., {Gadotti}, D.~A., {Elmegreen}, B.~G., {et~al.} 2020, in Galactic
  Dynamics in the Era of Large Surveys, ed. M.~{Valluri} \& J.~A. {Sellwood},
  Vol. 353, 140--143

\bibitem[{{Bland-Hawthorn} \& {Gerhard}(2016)}]{BlandHawthorn2016}
{Bland-Hawthorn}, J. \& {Gerhard}, O. 2016, \araa, 54, 529

\bibitem[{{Buta} \& {Block}(2001)}]{Buta2001}
{Buta}, R. \& {Block}, D.~L. 2001, \apj, 550, 243

\bibitem[{{Buta} {et~al.}(2001){Buta}, {Ryder}, {Madsen}, {Wesson}, {Crocker},
  \& {Combes}}]{buta01}
{Buta}, R., {Ryder}, S.~D., {Madsen}, G.~J., {et~al.} 2001, \aj, 121, 225

\bibitem[{{Buta}(2017)}]{Buta2017}
{Buta}, R.~J. 2017, \mnras, 470, 3819

\bibitem[{{Buta} {et~al.}(2015){Buta}, {Sheth}, {Athanassoula}, {Bosma},
  {Knapen}, {Laurikainen}, {Salo}, {Elmegreen}, {Ho}, \& {Zaritsky}}]{Buta2015}
{Buta}, R.~J., {Sheth}, K., {Athanassoula}, E., {et~al.} 2015, \apjs, 217, 32

\bibitem[{{Buta} \& {Zhang}(2009)}]{Buta2009}
{Buta}, R.~J. \& {Zhang}, X. 2009, \apjs, 182, 559

\bibitem[{{Combes} \& {Sanders}(1981)}]{Combes1981}
{Combes}, F. \& {Sanders}, R.~H. 1981, \aap, 96, 164

\bibitem[{{Comer{\'o}n} {et~al.}(2014){Comer{\'o}n}, {Salo}, {Laurikainen},
  {Knapen}, {Buta}, {Herrera-Endoqui}, {Laine}, {Holwerda}, {Sheth}, {Regan},
  {Hinz}, {Mu{\~n}oz-Mateos}, {Gil de Paz}, {Men{\'e}ndez-Delmestre},
  {Seibert}, {Mizusawa}, {Kim}, {Erroz-Ferrer}, {Gadotti}, {Athanassoula},
  {Bosma}, \& {Ho}}]{Comeron2014}
{Comer{\'o}n}, S., {Salo}, H., {Laurikainen}, E., {et~al.} 2014, \aap, 562,
  A121

\bibitem[{{Contopoulos}(1981)}]{Contopoulos1981}
{Contopoulos}, G. 1981, \aap, 102, 265

\bibitem[{{Contopoulos} \& {Grosbol}(1989)}]{Contopoulos1989}
{Contopoulos}, G. \& {Grosbol}, P. 1989, \aapr, 1, 261

\bibitem[{{Contopoulos} \& {Papayannopoulos}(1980)}]{contopoulos1980}
{Contopoulos}, G. \& {Papayannopoulos}, T. 1980, \aap, 92, 33

\bibitem[{{Corsini}(2011)}]{Corsini2011}
{Corsini}, E.~M. 2011, Mem. Soc. Astron. Ital. Suppl., 18, 23

\bibitem[{{Corsini} {et~al.}(2003){Corsini}, {Debattista}, \&
  {Aguerri}}]{Corsini2003}
{Corsini}, E.~M., {Debattista}, V.~P., \& {Aguerri}, J.~A.~L. 2003, \apjl, 599,
  L29

\bibitem[{{Cuomo} {et~al.}(2020){Cuomo}, {Aguerri}, {Corsini}, \&
  {Debattista}}]{Cuomo2020}
{Cuomo}, V., {Aguerri}, J.~A.~L., {Corsini}, E.~M., \& {Debattista}, V.~P.
  2020, \aap, 641, A111

\bibitem[{{Cuomo} {et~al.}(2019{\natexlab{a}}){Cuomo}, {Corsini}, {Aguerri},
  {Debattista}, {Coccato}, {Costantin}, {Dalla Bont{\`a}}, {Iodice},
  {M{\'e}ndez-Abreu}, {Morelli}, {Pagotto}, \& {Pizzella}}]{Cuomo2019}
{Cuomo}, V., {Corsini}, E.~M., {Aguerri}, J.~A.~L., {et~al.}
  2019{\natexlab{a}}, \mnras, 488, 4972

\bibitem[{{Cuomo} {et~al.}(2019{\natexlab{b}}){Cuomo}, {Lopez Aguerri},
  {Corsini}, {Debattista}, {M{\'e}ndez-Abreu}, \& {Pizzella}}]{Cuomo2019b}
{Cuomo}, V., {Lopez Aguerri}, J.~A., {Corsini}, E.~M., {et~al.}
  2019{\natexlab{b}}, \aap, 632, A51

\bibitem[{de~Vaucouleurs {et~al.}(1991)de~Vaucouleurs, {de Vaucouleurs},
  {Corwin}, {Buta}, {Paturel}, \& {Fouqu{\'e}}}]{deVaucouleurs1991}
de~Vaucouleurs, G., {de Vaucouleurs}, A., {Corwin}, Jr., H.~G., {et~al.} 1991,
  {Third Reference Catalogue of Bright Galaxies} (Springer-Verlag, New York
  USA)

\bibitem[{{Debattista}(2003)}]{Debattista2003}
{Debattista}, V.~P. 2003, \mnras, 342, 1194

\bibitem[{{Debattista} {et~al.}(2002{\natexlab{a}}){Debattista}, {Corsini}, \&
  {Aguerri}}]{Debattista2002}
{Debattista}, V.~P., {Corsini}, E.~M., \& {Aguerri}, J.~A.~L.
  2002{\natexlab{a}}, \mnras, 332, 65

\bibitem[{{Debattista} {et~al.}(2002{\natexlab{b}}){Debattista}, {Gerhard}, \&
  {Sevenster}}]{Debattista2002b}
{Debattista}, V.~P., {Gerhard}, O., \& {Sevenster}, M.~N. 2002{\natexlab{b}},
  \mnras, 334, 355

\bibitem[{{Debattista} \& {Sellwood}(1998)}]{Debattista1998}
{Debattista}, V.~P. \& {Sellwood}, J.~A. 1998, \apjl, 493, L5

\bibitem[{{Debattista} \& {Sellwood}(2000)}]{Debattista2000}
{Debattista}, V.~P. \& {Sellwood}, J.~A. 2000, \apj, 543, 704

\bibitem[{{D{\'\i}az-Garc{\'\i}a} {et~al.}(2016){D{\'\i}az-Garc{\'\i}a},
  {Salo}, \& {Laurikainen}}]{DiazGarcia2016}
{D{\'\i}az-Garc{\'\i}a}, S., {Salo}, H., \& {Laurikainen}, E. 2016, \aap, 596,
  A84

\bibitem[{{Elmegreen}(1996)}]{Elmegreen1996bis}
{Elmegreen}, B. 1996, in ASP Conf. Ser., Vol.~91, Barred Galaxies, ed.
  R.~{Buta}, D.~A. {Crocker}, \& B.~G. {Elmegreen} (Astron. Soc. Pac., San
  Francisco, CA), 197

\bibitem[{{Falc{\'o}n-Barroso} {et~al.}(2017){Falc{\'o}n-Barroso}, {Lyubenova},
  {van de Ven}, {Mendez-Abreu}, {Aguerri}, {Garc{\'{\i}}a-Lorenzo},
  {Bekerait{\'e}}, {S{\'a}nchez}, {Husemann}, {Garc{\'{\i}}a-Benito}, {Mast},
  {Walcher}, {Zibetti}, {Barrera-Ballesteros}, {Galbany},
  {S{\'a}nchez-Bl{\'a}zquez}, {Singh}, {van den Bosch}, {Wild}, {Zhu},
  {Bland-Hawthorn}, {Cid Fernandes}, {de Lorenzo-C{\'a}ceres}, {Gallazzi},
  {Gonz{\'a}lez Delgado}, {Marino}, {M{\'a}rquez}, {P{\'e}rez}, {P{\'e}rez},
  {Roth}, {Rosales-Ortega}, {Ruiz-Lara}, {Wisotzki}, {Ziegler}, \& {Califa
  Collaboration}}]{FalconBarroso2017}
{Falc{\'o}n-Barroso}, J., {Lyubenova}, M., {van de Ven}, G., {et~al.} 2017,
  \aap, 597, A48

\bibitem[{{Font} {et~al.}(2011){Font}, {Beckman}, {Epinat}, {Fathi},
  {Guti{\'e}rrez}, \& {Hernandez}}]{font2011}
{Font}, J., {Beckman}, J.~E., {Epinat}, B., {et~al.} 2011, \apjl, 741, L14

\bibitem[{{Friedli}(1999)}]{Friedli1999}
{Friedli}, D. 1999, in ASP Conf. Ser., Vol. 187, The Evolution of Galaxies on
  Cosmological Timescales, ed. J.~E. {Beckman} \& T.~J. {Mahoney} (Astron. Soc.
  Pac., San Francisco, CA), 88

\bibitem[{{Fuchs}(2001)}]{Fuchs2001}
{Fuchs}, B. 2001, in Dark Matter in Astro- and Particle Physics, ed. H.~V.
  {Klapdor-Kleingrothaus}, 25

\bibitem[{{Garcia-Lorenzo} {et~al.}(2015){Garcia-Lorenzo}, {Marquez},
  {Barrera-Ballesteros}, {Masegosa}, {Husemann}, {Falcon-Barroso}, {Lyubenova},
  {Sanchez}, {Walcher}, {Mast}, {Garcia-Benito}, {Mendez-Abreu}, {Van De Ven},
  {Spekkens}, {Holmes}, {Monreal-Ibero}, {Del Olmo}, {Ziegler},
  {Bland-Hawthorn}, {Sanchez-Blazquez}, {Iglesias-Paramo}, {Aguerri},
  {Papaderos}, {Gomes}, {Marino}, {Gonzalez Delgado}, {Cortijo-Ferrero},
  {Lopez-Sanchez}, {Bekeraite}, {Wisotzki}, {Bomans}, \& {CALIFA
  Team}}]{Garcia-Lorenzo2015}
{Garcia-Lorenzo}, B., {Marquez}, I., {Barrera-Ballesteros}, J.~K., {et~al.}
  2015, VizieR Online Data Catalog, J/A+A/573/A59

\bibitem[{{Garma-Oehmichen} {et~al.}(2020){Garma-Oehmichen}, {Cano-D{\'\i}az},
  {Hern{\'a}ndez-Toledo}, {Aquino-Ort{\'\i}z}, {Valenzuela}, {Aguerri},
  {S{\'a}nchez}, \& {Merrifield}}]{Garma-Oehmichen2020}
{Garma-Oehmichen}, L., {Cano-D{\'\i}az}, M., {Hern{\'a}ndez-Toledo}, H.,
  {et~al.} 2020, \mnras, 491, 3655

\bibitem[{{Gerssen} \& {Debattista}(2007)}]{gerssen2007}
{Gerssen}, J. \& {Debattista}, V.~P. 2007, \mnras, 378, 189

\bibitem[{{Ghafourian} {et~al.}(2020){Ghafourian}, {Roshan}, \&
  {Abbassi}}]{Ghafourian2020}
{Ghafourian}, N., {Roshan}, M., \& {Abbassi}, S. 2020, \apj, 895, 13

\bibitem[{{Guo} {et~al.}(2019){Guo}, {Mao}, {Athanassoula}, {Li}, {Ge}, {Long},
  {Merrifield}, \& {Masters}}]{Guo2019}
{Guo}, R., {Mao}, S., {Athanassoula}, E., {et~al.} 2019, \mnras, 482, 1733

\bibitem[{{Hernquist} \& {Weinberg}(1992)}]{Hernquist1992}
{Hernquist}, L. \& {Weinberg}, M.~D. 1992, \apj, 400, 80

\bibitem[{{Hilmi} {et~al.}(2020){Hilmi}, {Minchev}, {Buck}, {Martig},
  {Quillen}, {Monari}, {Famaey}, {de Jong}, {Laporte}, {Read}, {Sand ers},
  {Steinmetz}, \& {Wegg}}]{Hilmi2020}
{Hilmi}, T., {Minchev}, I., {Buck}, T., {et~al.} 2020, \mnras, 497, 933

\bibitem[{{Hinshaw} {et~al.}(2013){Hinshaw}, {Larson}, {Komatsu}, {Spergel},
  {Bennett}, {Dunkley}, {Nolta}, {Halpern}, {Hill}, {Odegard}, {Page}, {Smith},
  {Weiland}, {Gold}, {Jarosik}, {Kogut}, {Limon}, {Meyer}, {Tucker}, {Wollack},
  \& {Wright}}]{Hinshaw2013}
{Hinshaw}, G., {Larson}, D., {Komatsu}, E., {et~al.} 2013, \apjs, 208, 19

\bibitem[{{James} \& {Percival}(2016)}]{James2016}
{James}, P.~A. \& {Percival}, S.~M. 2016, \mnras, 457, 917

\bibitem[{{Jedrzejewski}(1987)}]{Jedrzejewski1987}
{Jedrzejewski}, R.~I. 1987, \mnras, 226, 747

\bibitem[{{Kim} {et~al.}(2016){Kim}, {Gadotti}, {Athanassoula}, {Bosma},
  {Sheth}, \& {Lee}}]{Kim2016}
{Kim}, T., {Gadotti}, D.~A., {Athanassoula}, E., {et~al.} 2016, \mnras, 462,
  3430

\bibitem[{{Kormendy}(1979)}]{Kormendy1979}
{Kormendy}, J. 1979, \apj, 227, 714

\bibitem[{{Laurikainen} \& {Salo}(2002)}]{Laurikainen2002}
{Laurikainen}, E. \& {Salo}, H. 2002, \mnras, 337, 1118

\bibitem[{{Lee} {et~al.}(2019){Lee}, {Ann}, \& {Park}}]{Lee2019}
{Lee}, Y.~H., {Ann}, H.~B., \& {Park}, M.-G. 2019, \apj, 872, 97

\bibitem[{{Lee} {et~al.}(2020){Lee}, {Park}, {Ann}, {Kim}, \& {Seo}}]{Lee2020}
{Lee}, Y.~H., {Park}, M.-G., {Ann}, H.~B., {Kim}, T., \& {Seo}, W.-Y. 2020,
  \apj, 899, 84

\bibitem[{{Lindblad} \& {Kristen}(1996)}]{lindblad1996}
{Lindblad}, P.~A.~B. \& {Kristen}, H. 1996, \aap, 313, 733

\bibitem[{{Little} \& {Carlberg}(1991)}]{little1991}
{Little}, B. \& {Carlberg}, R.~G. 1991, \mnras, 251, 227

\bibitem[{{Maciejewski}(2006)}]{Maciejewski2006}
{Maciejewski}, W. 2006, \mnras, 371, 451

\bibitem[{{Marinova} \& {Jogee}(2007)}]{Marinova2007}
{Marinova}, I. \& {Jogee}, S. 2007, \apj, 659, 1176

\bibitem[{{Martinez-Valpuesta} {et~al.}(2017){Martinez-Valpuesta}, {Aguerri},
  {Gonz{\'a}lez-Garc{\'\i}a}, {Dalla Vecchia}, \&
  {Stringer}}]{MartinezValpuesta2017}
{Martinez-Valpuesta}, I., {Aguerri}, J. A.~L., {Gonz{\'a}lez-Garc{\'\i}a},
  A.~C., {Dalla Vecchia}, C., \& {Stringer}, M. 2017, \mnras, 464, 1502

\bibitem[{{Meidt} {et~al.}(2009){Meidt}, {Rand}, \& {Merrifield}}]{Meidt2009}
{Meidt}, S.~E., {Rand}, R.~J., \& {Merrifield}, M.~R. 2009, \apj, 702, 277

\bibitem[{{Meidt} {et~al.}(2008){Meidt}, {Rand}, {Merrifield}, {Debattista}, \&
  {Shen}}]{Meidt2008}
{Meidt}, S.~E., {Rand}, R.~J., {Merrifield}, M.~R., {Debattista}, V.~P., \&
  {Shen}, J. 2008, \apj, 676, 899

\bibitem[{{M{\'e}ndez-Abreu} {et~al.}(2017){M{\'e}ndez-Abreu}, {Ruiz-Lara},
  {S{\'a}nchez-Menguiano}, {de Lorenzo-C{\'a}ceres}, {Costantin},
  {Catal{\'a}n-Torrecilla}, {Florido}, {Aguerri}, {Bland-Hawthorn}, {Corsini},
  {Dettmar}, {Galbany}, {Garc{\'{\i}}a-Benito}, {Marino}, {M{\'a}rquez},
  {Ortega-Minakata}, {Papaderos}, {S{\'a}nchez}, {S{\'a}nchez-Blazquez},
  {Spekkens}, {van de Ven}, {Wild}, \& {Ziegler}}]{MendezAbreu2017}
{M{\'e}ndez-Abreu}, J., {Ruiz-Lara}, T., {S{\'a}nchez-Menguiano}, L., {et~al.}
  2017, \aap, 598, A32

\bibitem[{{Men{\'e}ndez-Delmestre} {et~al.}(2007){Men{\'e}ndez-Delmestre},
  {Sheth}, {Schinnerer}, {Jarrett}, \& {Scoville}}]{MenendezDelmestre2007}
{Men{\'e}ndez-Delmestre}, K., {Sheth}, K., {Schinnerer}, E., {Jarrett}, T.~H.,
  \& {Scoville}, N.~Z. 2007, \apj, 657, 790

\bibitem[{{Merrifield} \& {Kuijken}(1995)}]{Merrifield1995}
{Merrifield}, M.~R. \& {Kuijken}, K. 1995, \mnras, 274, 933

\bibitem[{{Michel-Dansac} \& {Wozniak}(2006)}]{MichelDansac2006}
{Michel-Dansac}, L. \& {Wozniak}, H. 2006, \aap, 452, 97

\bibitem[{{Morelli} {et~al.}(2016){Morelli}, {Parmiggiani}, {Corsini},
  {Costantin}, {Dalla Bont{\`a}}, {M{\'e}ndez-Abreu}, \&
  {Pizzella}}]{morelli2016}
{Morelli}, L., {Parmiggiani}, M., {Corsini}, E.~M., {et~al.} 2016, \mnras, 463,
  4396

\bibitem[{{Noguchi}(1988)}]{Noguchi1988}
{Noguchi}, M. 1988, \aap, 203, 259

\bibitem[{{Ohta} {et~al.}(1990){Ohta}, {Hamabe}, \& {Wakamatsu}}]{Ohta1990}
{Ohta}, K., {Hamabe}, M., \& {Wakamatsu}, K.-I. 1990, \apj, 357, 71

\bibitem[{{O'Neill} \& {Dubinski}(2003)}]{oneill2003}
{O'Neill}, J.~K. \& {Dubinski}, J. 2003, \mnras, 346, 251

\bibitem[{{Palunas} \& {Williams}(2000)}]{Palunas2000}
{Palunas}, P. \& {Williams}, T.~B. 2000, \aj, 120, 2884

\bibitem[{{Petersen} {et~al.}(2019){Petersen}, {Weinberg}, \&
  {Katz}}]{Petersen2019}
{Petersen}, M.~S., {Weinberg}, M.~D., \& {Katz}, N. 2019, \mnras, 490, 3616

\bibitem[{{Puerari} \& {Dottori}(1997)}]{puerari1997}
{Puerari}, I. \& {Dottori}, H. 1997, \apjl, 476, L73

\bibitem[{{Quillen} {et~al.}(1994){Quillen}, {Frogel}, \&
  {Gonzalez}}]{Quillen1994}
{Quillen}, A.~C., {Frogel}, J.~A., \& {Gonzalez}, R.~A. 1994, \apj, 437, 162

\bibitem[{{Rautiainen} {et~al.}(2008){Rautiainen}, {Salo}, \&
  {Laurikainen}}]{Rautiainen2008}
{Rautiainen}, P., {Salo}, H., \& {Laurikainen}, E. 2008, \mnras, 388, 1803

\bibitem[{{Reyes} {et~al.}(2011){Reyes}, {Mandelbaum}, {Gunn}, {Pizagno}, \&
  {Lackner}}]{Reyes2011}
{Reyes}, R., {Mandelbaum}, R., {Gunn}, J.~E., {Pizagno}, J., \& {Lackner},
  C.~N. 2011, \mnras, 417, 2347

\bibitem[{{S{\'a}nchez} {et~al.}(2012){S{\'a}nchez}, {Kennicutt}, {Gil de Paz},
  {van de Ven}, {V{\'{\i}}lchez}, {Wisotzki}, {Walcher}, {Mast}, {Aguerri},
  {Albiol-P{\'e}rez}, {Alonso-Herrero}, {Alves}, {Bakos}, {Bart{\'a}kov{\'a}},
  {Bland-Hawthorn}, {Boselli}, {Bomans}, {Castillo-Morales}, {Cortijo-Ferrero},
  {de Lorenzo-C{\'a}ceres}, {Del Olmo}, {Dettmar}, {D{\'{\i}}az}, {Ellis},
  {Falc{\'o}n-Barroso}, {Flores}, {Gallazzi}, {Garc{\'{\i}}a-Lorenzo},
  {Gonz{\'a}lez Delgado}, {Gruel}, {Haines}, {Hao}, {Husemann},
  {Igl{\'e}sias-P{\'a}ramo}, {Jahnke}, {Johnson}, {Jungwiert}, {Kalinova},
  {Kehrig}, {Kupko}, {L{\'o}pez-S{\'a}nchez}, {Lyubenova}, {Marino},
  {M{\'a}rmol-Queralt{\'o}}, {M{\'a}rquez}, {Masegosa}, {Meidt},
  {Mendez-Abreu}, {Monreal-Ibero}, {Montijo}, {Mour{\~a}o}, {Palacios-Navarro},
  {Papaderos}, {Pasquali}, {Peletier}, {P{\'e}rez}, {P{\'e}rez}, {Quirrenbach},
  {Rela{\~n}o}, {Rosales-Ortega}, {Roth}, {Ruiz-Lara},
  {S{\'a}nchez-Bl{\'a}zquez}, {Sengupta}, {Singh}, {Stanishev}, {Trager},
  {Vazdekis}, {Viironen}, {Wild}, {Zibetti}, \& {Ziegler}}]{sanchez2012}
{S{\'a}nchez}, S.~F., {Kennicutt}, R.~C., {Gil de Paz}, A., {et~al.} 2012,
  \aap, 538, A8

\bibitem[{{Sanders} \& {Huntley}(1976)}]{Sanders1976}
{Sanders}, R.~H. \& {Huntley}, J.~M. 1976, \apj, 209, 53

\bibitem[{{Sanders} \& {Tubbs}(1980)}]{Sanders1980}
{Sanders}, R.~H. \& {Tubbs}, A.~D. 1980, \apj, 235, 803

\bibitem[{{Sellwood}(1981)}]{Sellwood1981}
{Sellwood}, J.~A. 1981, \aap, 99, 362

\bibitem[{{Sellwood}(2014)}]{Sellwood2014}
{Sellwood}, J.~A. 2014, Reviews of Modern Physics, 86, 1

\bibitem[{{Sellwood} \& {Sparke}(1988)}]{Sellwood1988}
{Sellwood}, J.~A. \& {Sparke}, L.~S. 1988, \mnras, 231, 25P

\bibitem[{{Starkman} {et~al.}(2018){Starkman}, {Lelli}, {McGaugh}, \&
  {Schombert}}]{Starkman2018}
{Starkman}, N., {Lelli}, F., {McGaugh}, S., \& {Schombert}, J. 2018, \mnras,
  480, 2292

\bibitem[{{Theureau} {et~al.}(1998){Theureau}, {Bottinelli}, {Coudreau-Durand},
  {Gouguenheim}, {Hallet}, {Loulergue}, {Paturel}, \&
  {Teerikorpi}}]{Theureau1998}
{Theureau}, G., {Bottinelli}, L., {Coudreau-Durand}, N., {et~al.} 1998, \aaps,
  130, 333

\bibitem[{{Toomre}(1981)}]{Toomre1981}
{Toomre}, A. 1981, in Structure and Evolution of Normal Galaxies, ed. S.~M.
  {Fall} \& D.~{Lynden-Bell}, Proc. of the Advanced Study Institute (Cambridge
  University Press, Cambridge, UK), 111

\bibitem[{{Tremaine} \& {Weinberg}(1984)}]{Tremaine1984}
{Tremaine}, S. \& {Weinberg}, M.~D. 1984, \apjl, 282, L5

\bibitem[{{Tully} \& {Fisher}(1977)}]{Tully1977}
{Tully}, R.~B. \& {Fisher}, J.~R. 1977, \aap, 500, 105

\bibitem[{{Vasiliev} \& {Athanassoula}(2015)}]{Vasiliev2015}
{Vasiliev}, E. \& {Athanassoula}, E. 2015, \mnras, 450, 2842

\bibitem[{{Villa-Vargas} {et~al.}(2010){Villa-Vargas}, {Shlosman}, \&
  {Heller}}]{villavargas2010}
{Villa-Vargas}, J., {Shlosman}, I., \& {Heller}, C. 2010, \apj, 719, 1470

\bibitem[{{Walcher} {et~al.}(2014){Walcher}, {Wisotzki}, {Bekerait{\'e}},
  {Husemann}, {Iglesias-P{\'a}ramo}, {Backsmann}, {Barrera Ballesteros},
  {Catal{\'a}n-Torrecilla}, {Cortijo}, {del Olmo}, {Garcia Lorenzo},
  {Falc{\'o}n-Barroso}, {Jilkova}, {Kalinova}, {Mast}, {Marino},
  {M{\'e}ndez-Abreu}, {Pasquali}, {S{\'a}nchez}, {Trager}, {Zibetti},
  {Aguerri}, {Alves}, {Bland-Hawthorn}, {Boselli}, {Castillo Morales}, {Cid
  Fernandes}, {Flores}, {Galbany}, {Gallazzi}, {Garc{\'{\i}}a-Benito}, {Gil de
  Paz}, {Gonz{\'a}lez-Delgado}, {Jahnke}, {Jungwiert}, {Kehrig}, {Lyubenova},
  {M{\'a}rquez Perez}, {Masegosa}, {Monreal Ibero}, {P{\'e}rez}, {Quirrenbach},
  {Rosales-Ortega}, {Roth}, {Sanchez-Blazquez}, {Spekkens}, {Tundo}, {van de
  Ven}, {Verheijen}, {Vilchez}, \& {Ziegler}}]{Walcher2014}
{Walcher}, C.~J., {Wisotzki}, L., {Bekerait{\'e}}, S., {et~al.} 2014, \aap,
  569, A1

\bibitem[{{Weinberg}(1985)}]{weinberg1985}
{Weinberg}, M.~D. 1985, \mnras, 213, 451

\bibitem[{{Williams} {et~al.}(2021){Williams}, {Schinnerer}, {Emsellem},
  {Meidt}, {Querejeta}, {Belfiore}, {Be{\v{s}}li{\'c}}, {Bigiel}, {Chevance},
  {Dale}, {Glover}, {Grasha}, {Klessen}, {Kruijssen}, {Leroy}, {Pan}, {Pety},
  {Pessa}, {Rosolowsky}, {Saito}, {Santoro}, {Schruba}, {Sormani}, {Sun}, \&
  {Watkins}}]{Williams2021}
{Williams}, T.~G., {Schinnerer}, E., {Emsellem}, E., {et~al.} 2021, arXiv
  e-prints, arXiv:2102.01091

\bibitem[{{Wozniak} \& {Pierce}(1991)}]{Wozniak1991}
{Wozniak}, H. \& {Pierce}, M.~J. 1991, \aaps, 88, 325

\bibitem[{{Zhang} \& {Buta}(2007)}]{Zhang2007}
{Zhang}, X. \& {Buta}, R.~J. 2007, \aj, 133, 2584

\bibitem[{{Zou} {et~al.}(2019){Zou}, {Shen}, {Bureau}, \& {Li}}]{Zou2019}
{Zou}, Y., {Shen}, J., {Bureau}, M., \& {Li}, Z.-Y. 2019, \apj, 884, 23

\end{thebibliography}

\begin{appendix}

\section{Analysis of the ratio maps for the entire sample}
\label{app2}

We present here the analysis of the ratio maps for the entire sample.

\begin{figure*}
    \centering
    \includegraphics[scale=0.8]{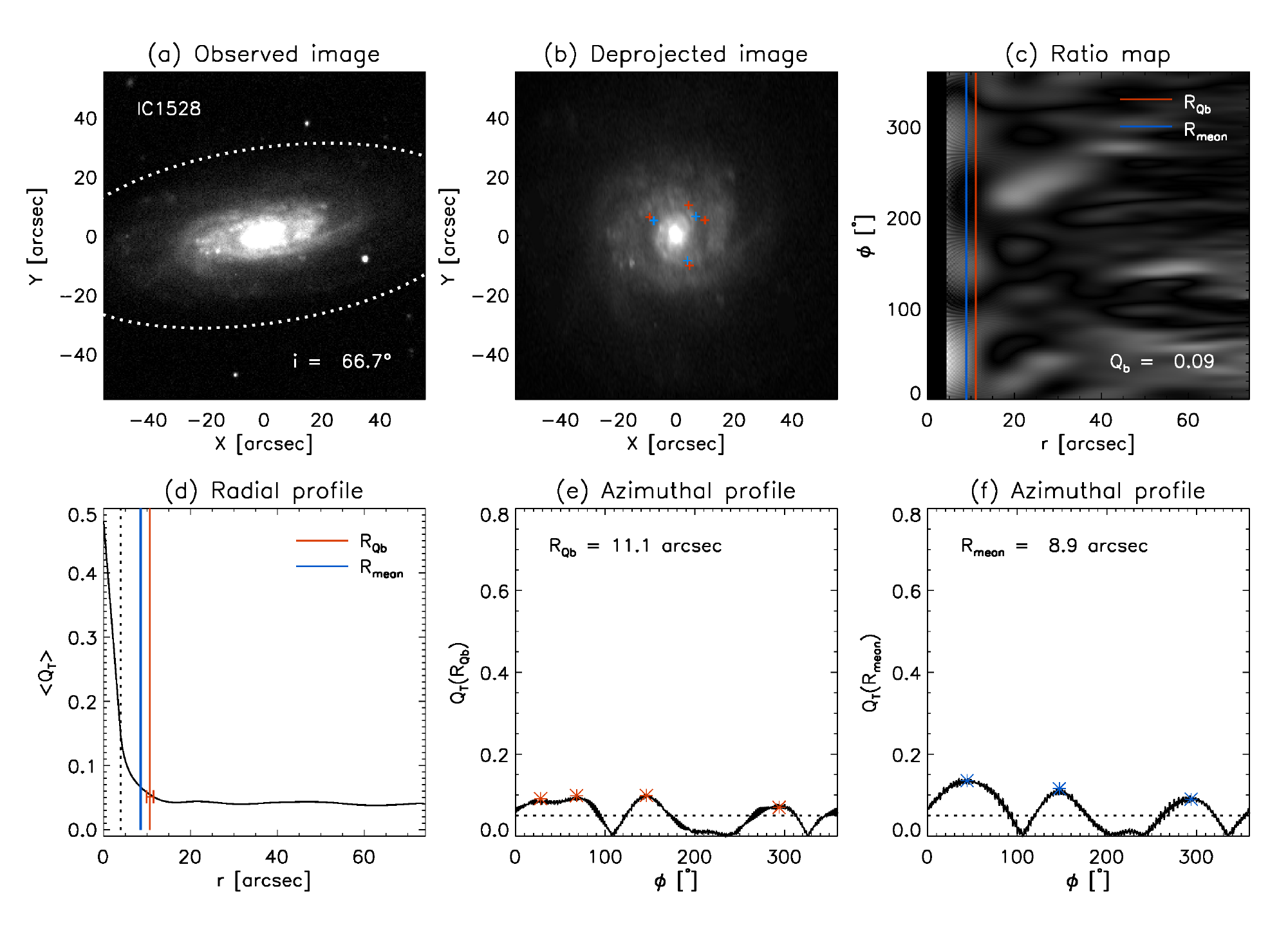}
    \includegraphics[scale=0.8]{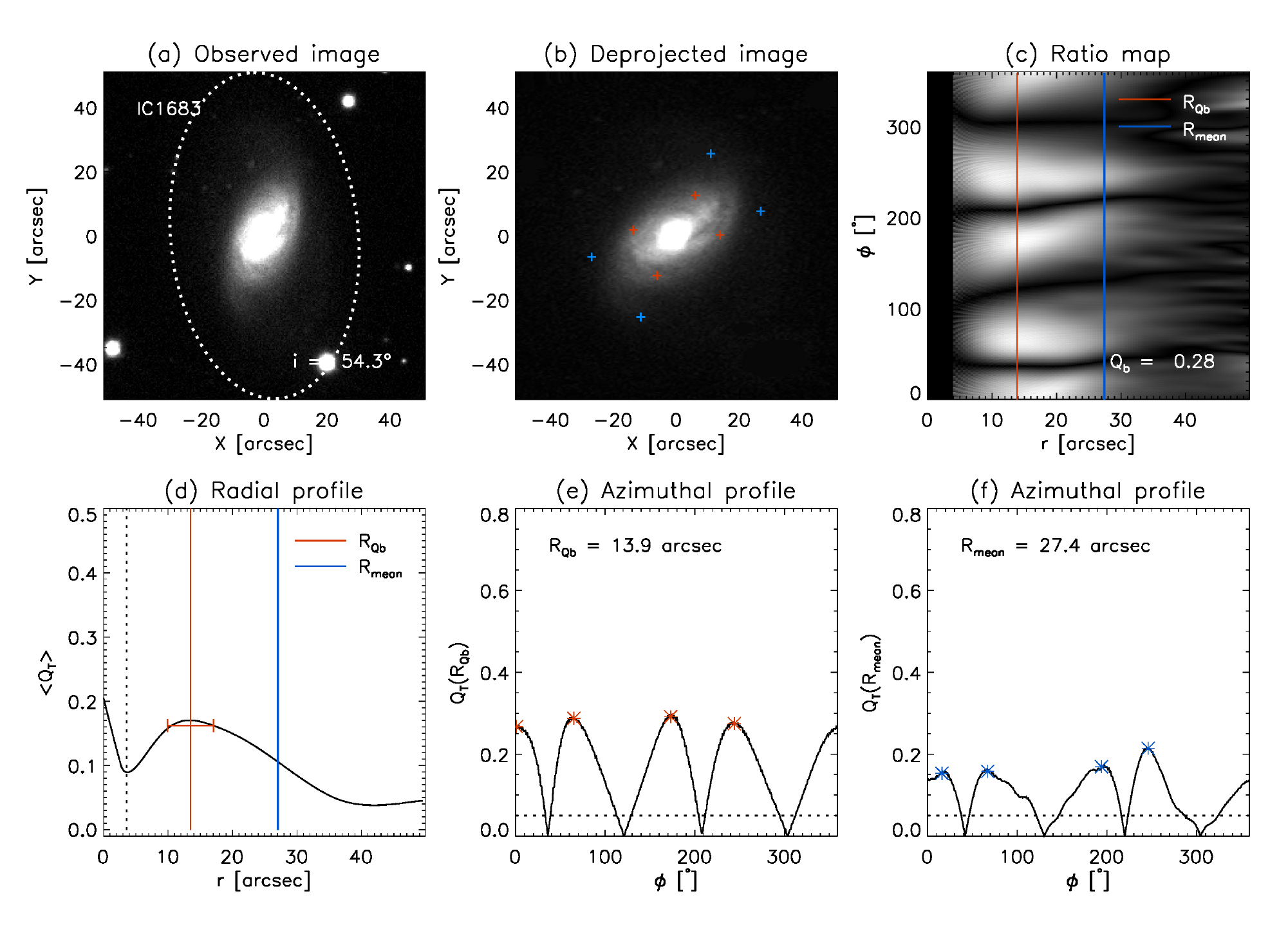}
    \caption{Same as Fig.~\ref{fig:pt_mapN5406} but for the remaining galaxies of the sample.}
    \label{fig:pt_map1}
\end{figure*}

\begin{figure*}\ContinuedFloat
    \centering
    \includegraphics[scale=0.8]{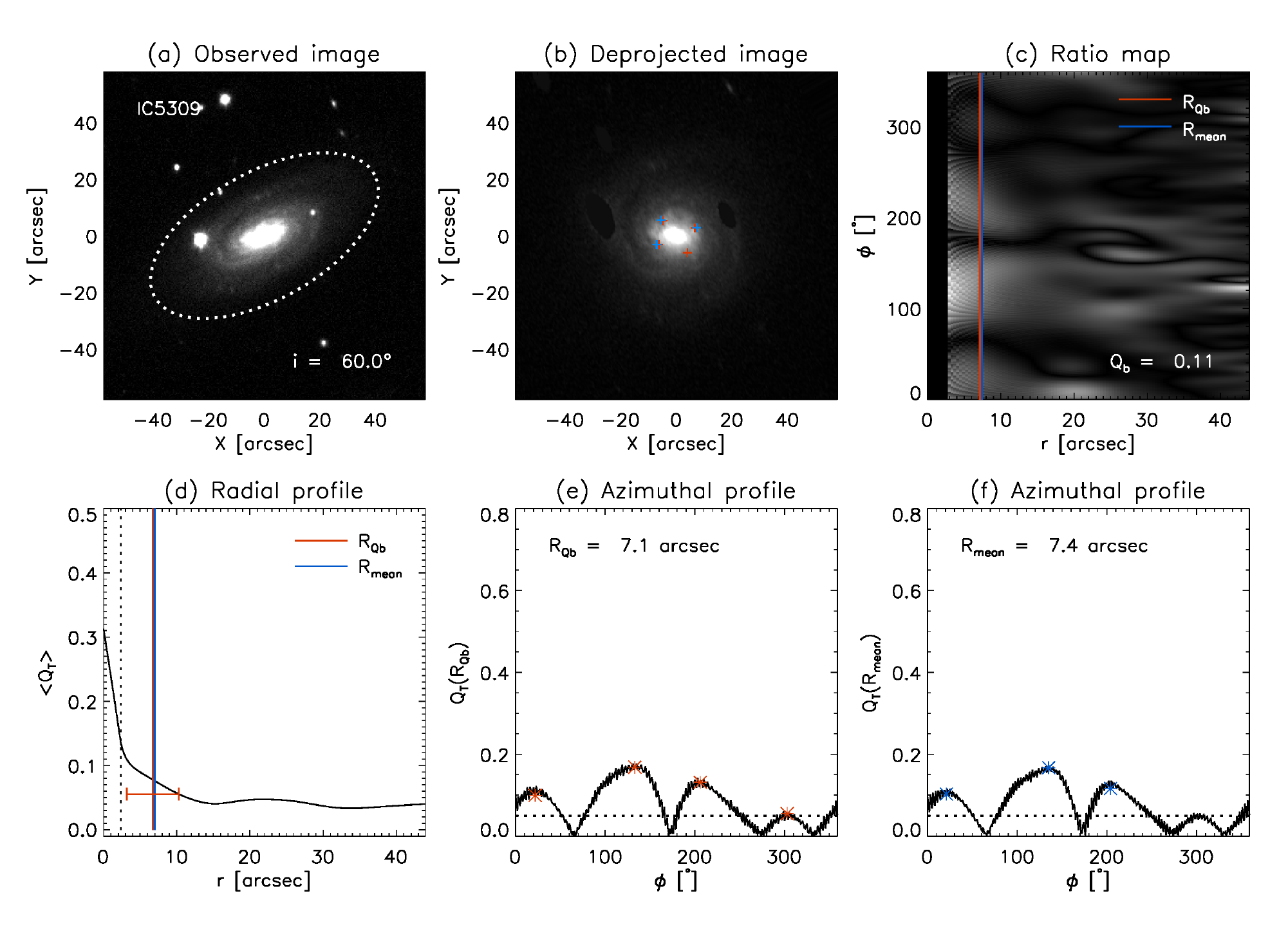}
    \includegraphics[scale=0.8]{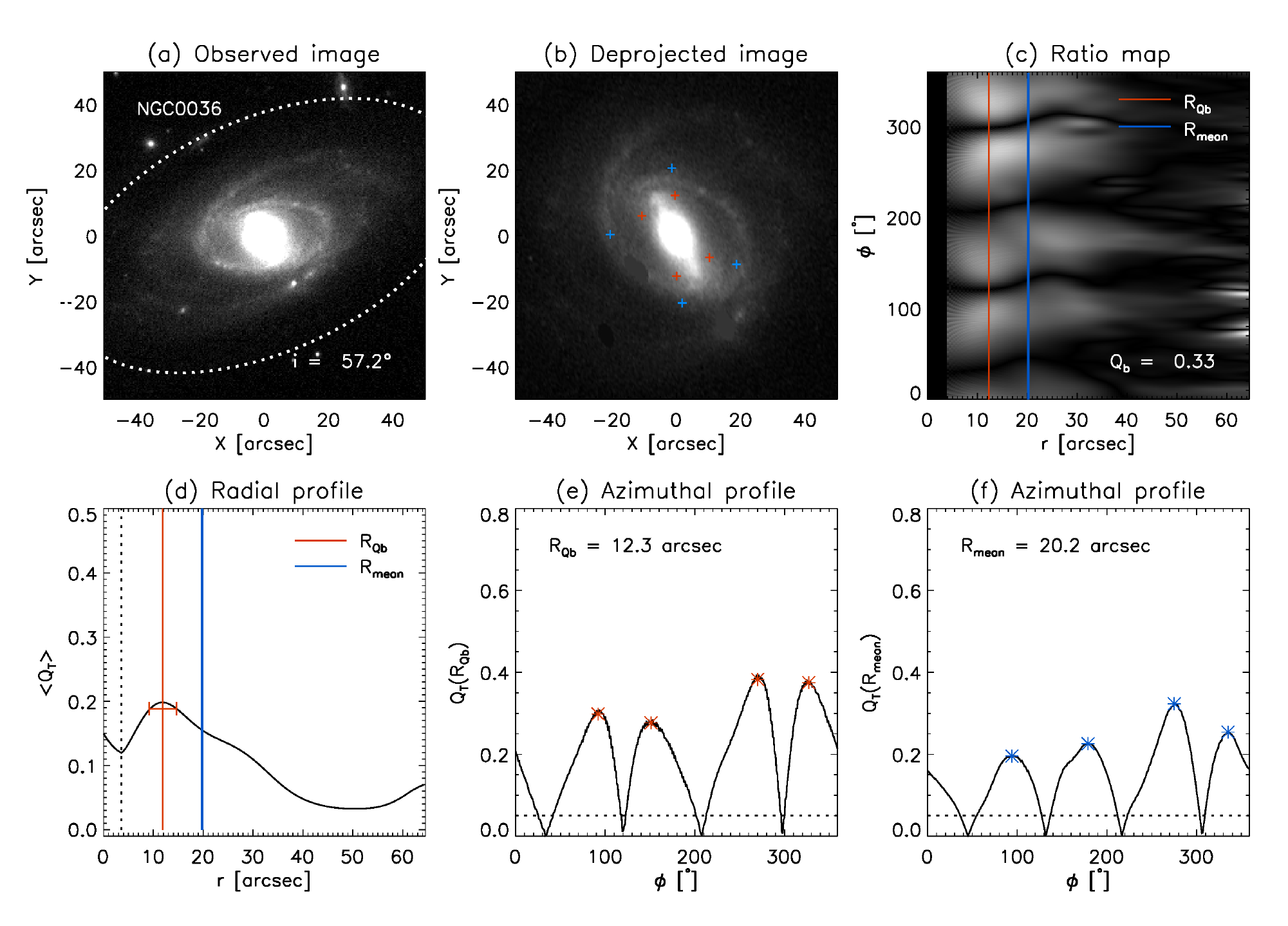}
    \caption{(continued).}
    \label{fig:pt_map3}
\end{figure*}

\begin{figure*}\ContinuedFloat
    \centering
    \includegraphics[scale=0.8]{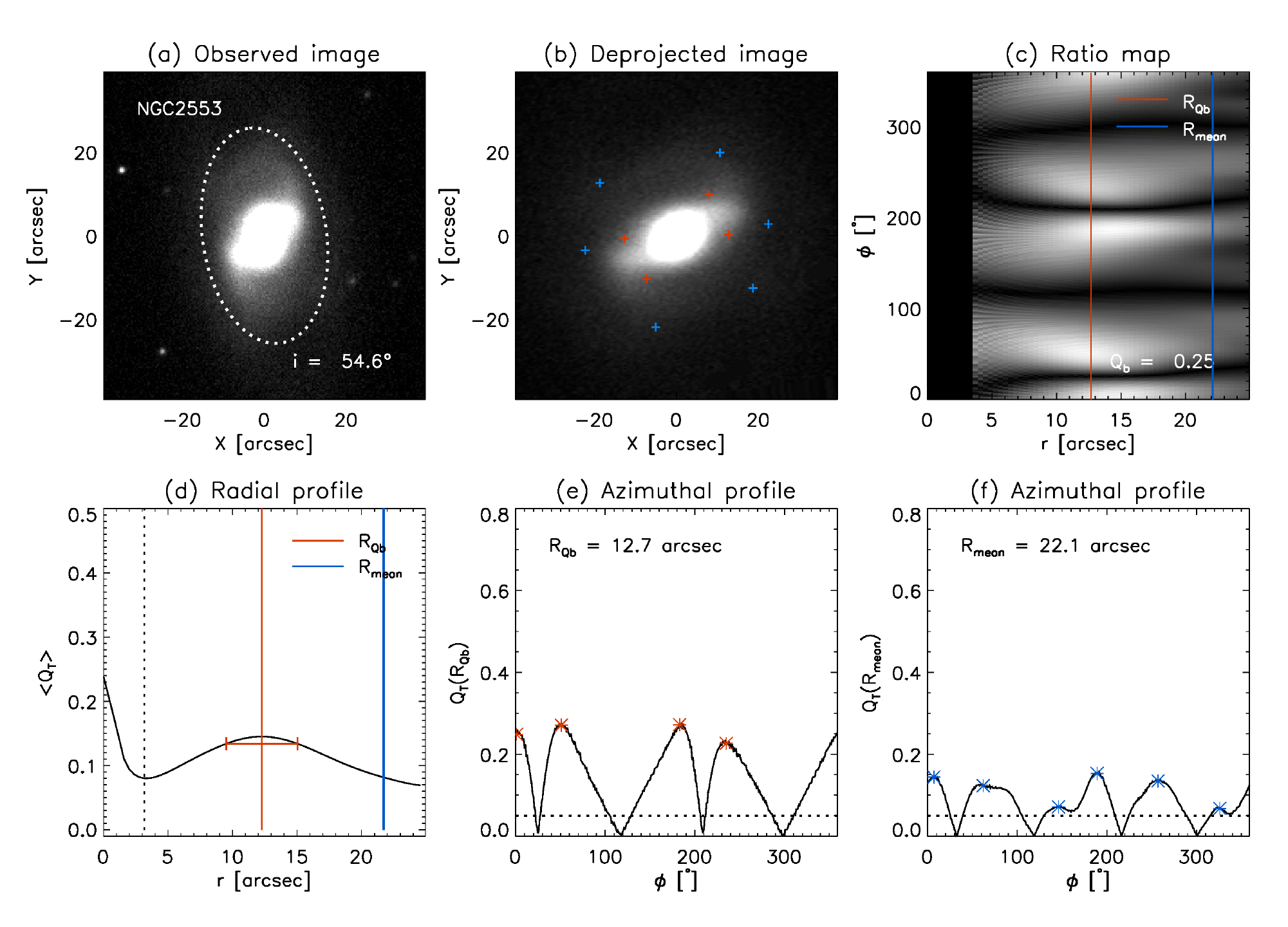}
    \includegraphics[scale=0.8]{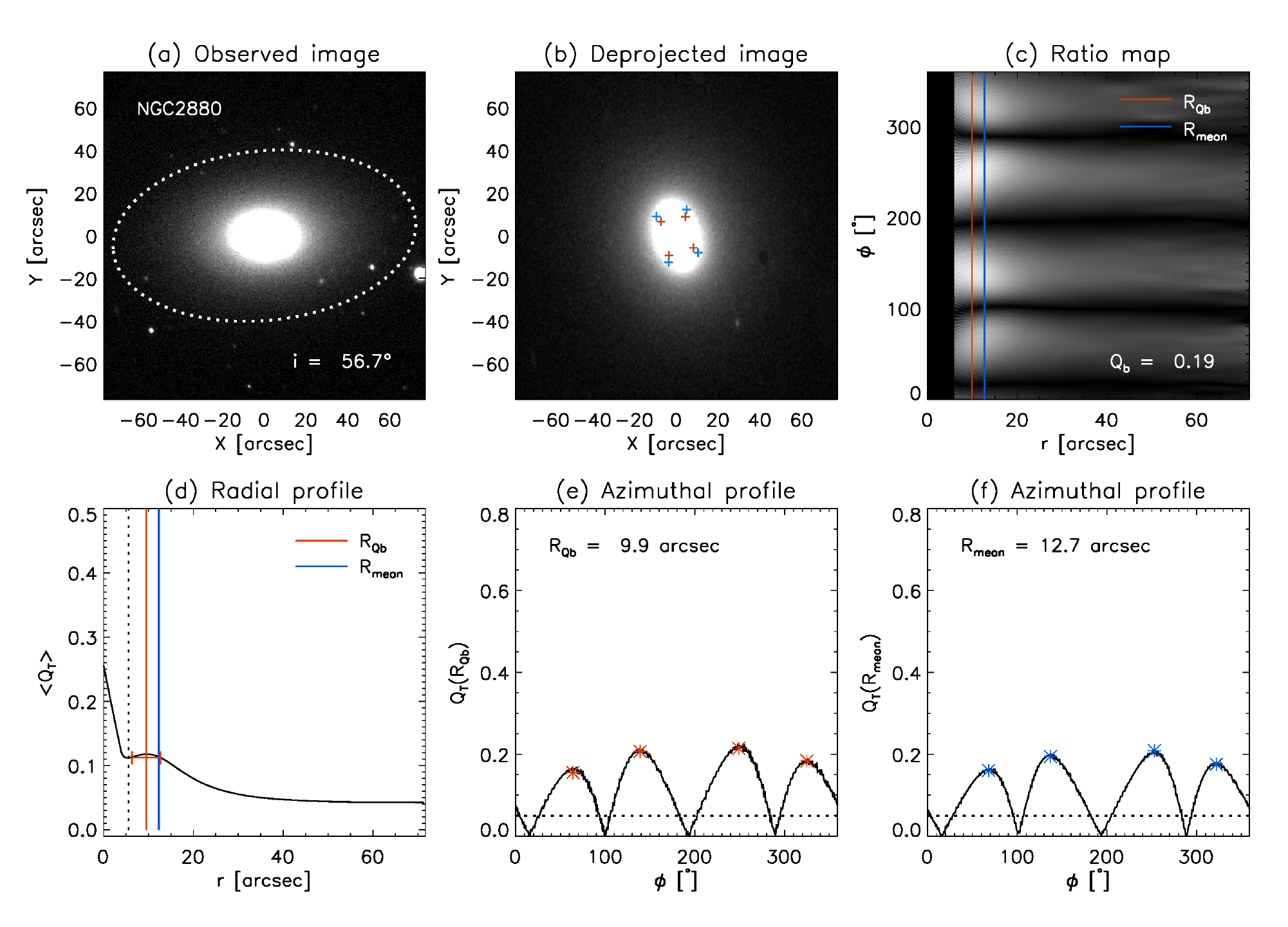}
    \caption{(continued).}
    \label{fig:pt_map5}
\end{figure*}

\begin{figure*}\ContinuedFloat
    \centering
    \includegraphics[scale=0.8]{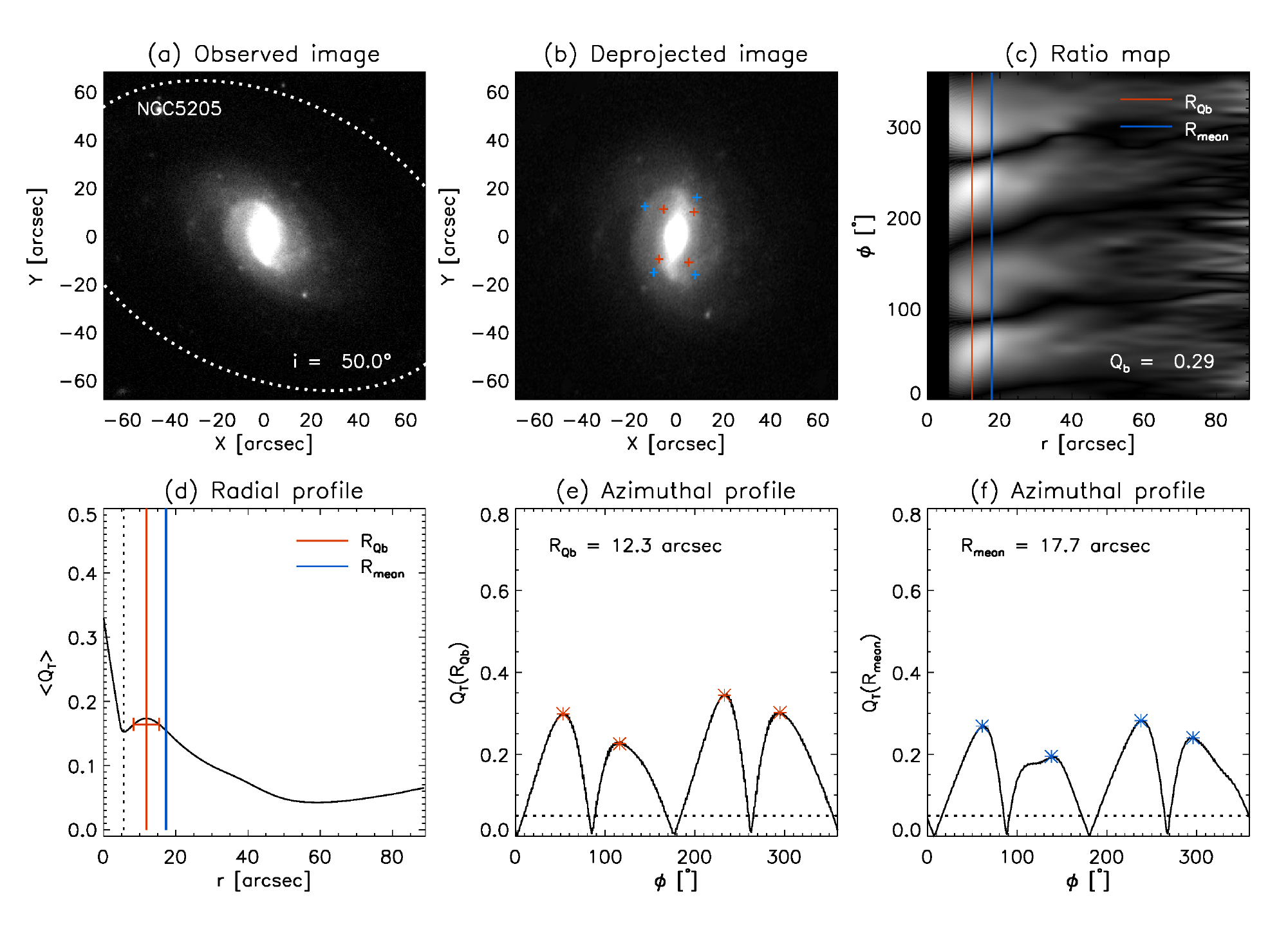}
    \includegraphics[scale=0.8]{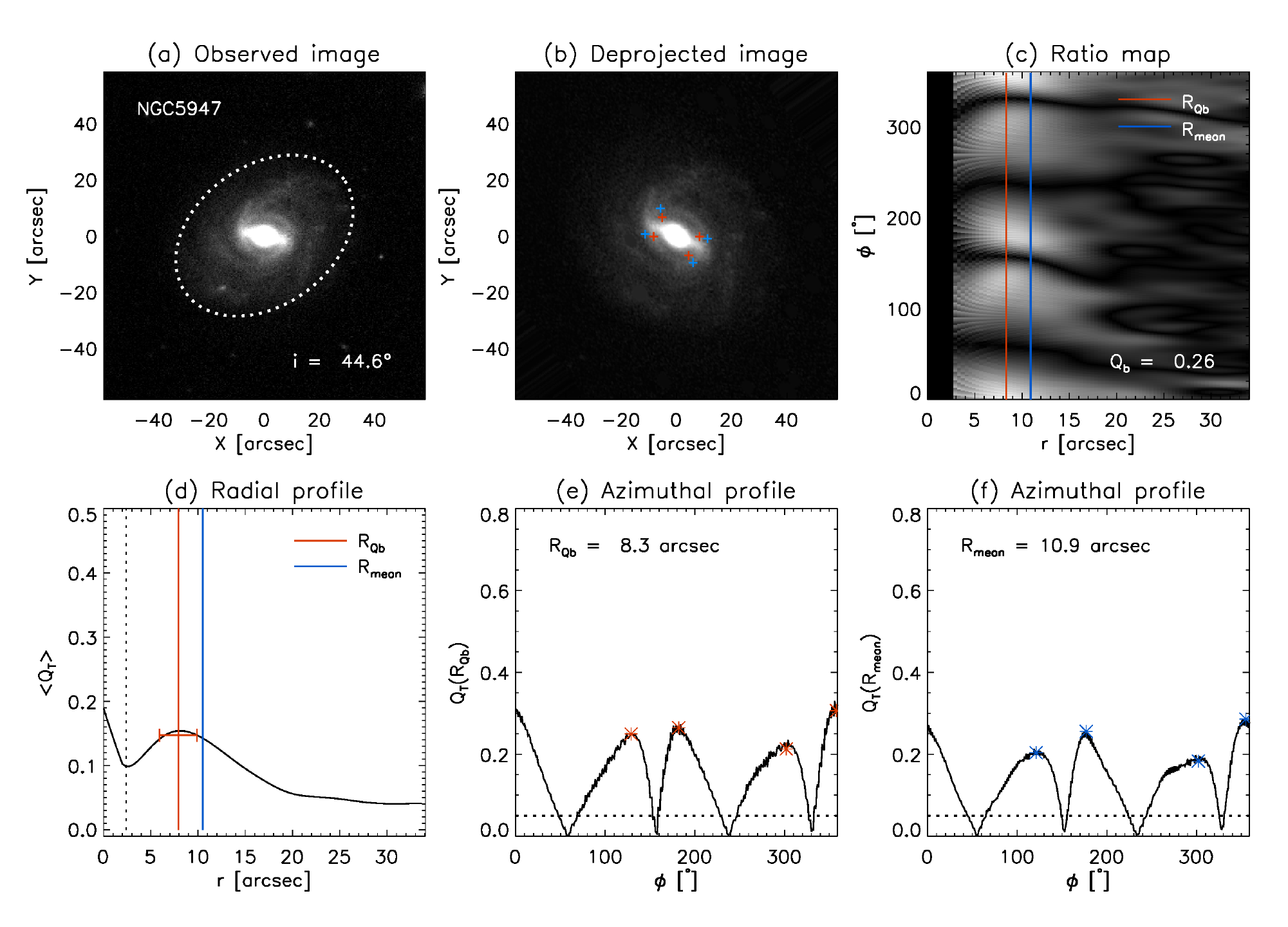}
    \caption{(continued).}
    \label{fig:pt_map6}
\end{figure*}

\begin{figure*}\ContinuedFloat
    \centering
    \includegraphics[scale=0.8]{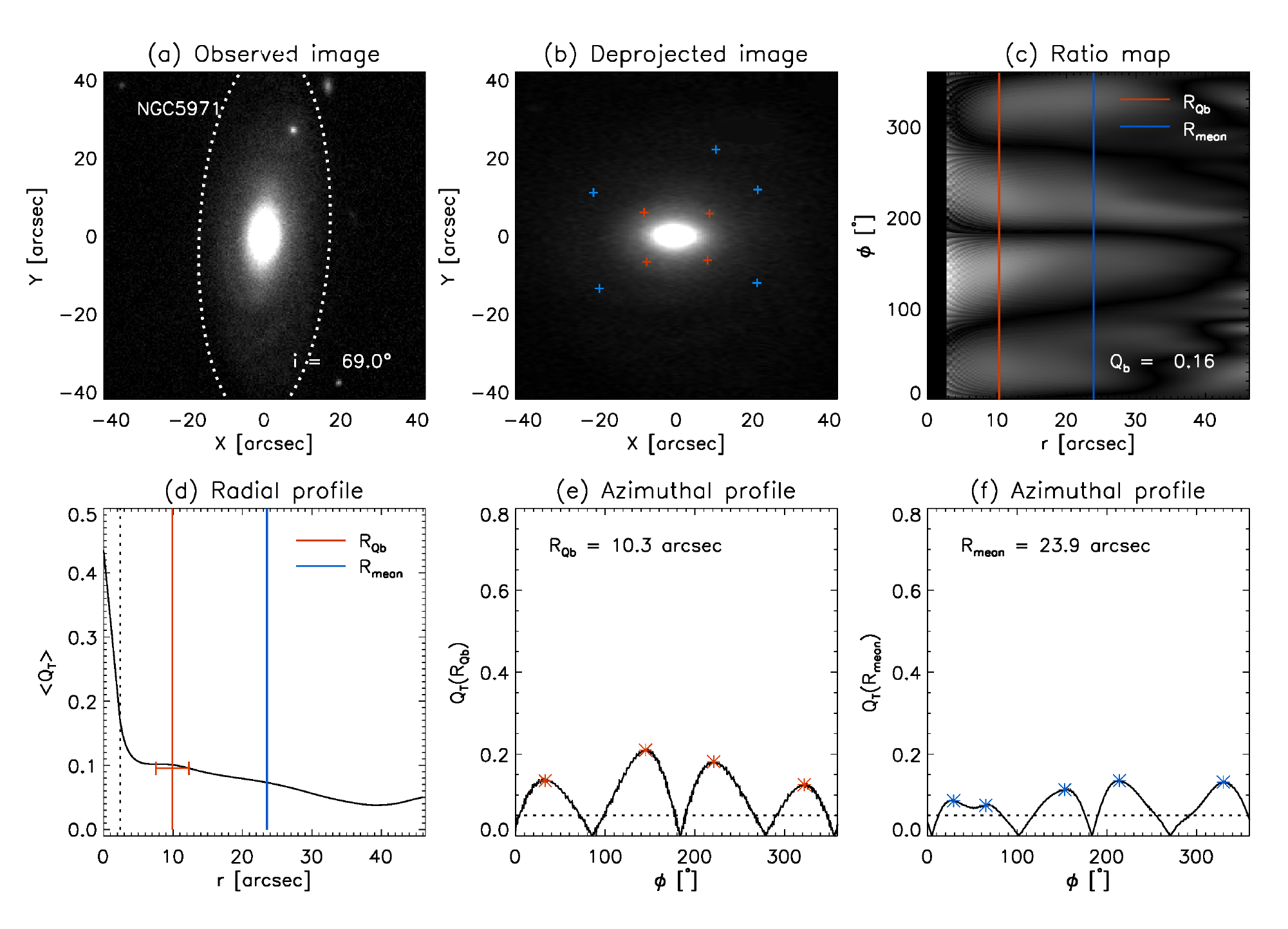}
    \includegraphics[scale=0.8]{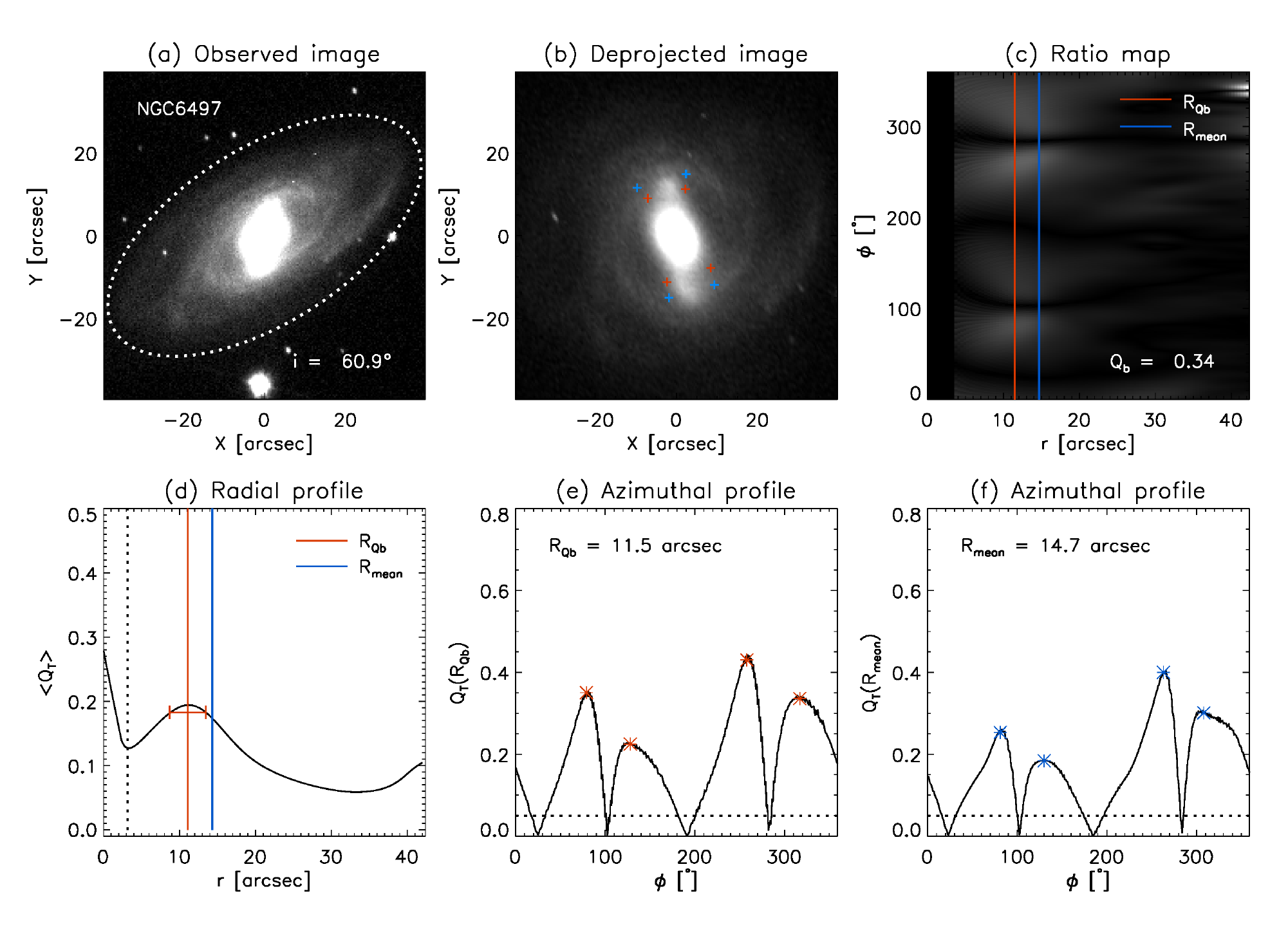}
    \caption{(continued).}
    \label{fig:pt_map6}
\end{figure*}

\begin{figure*}\ContinuedFloat
    \centering
    \includegraphics[scale=0.8]{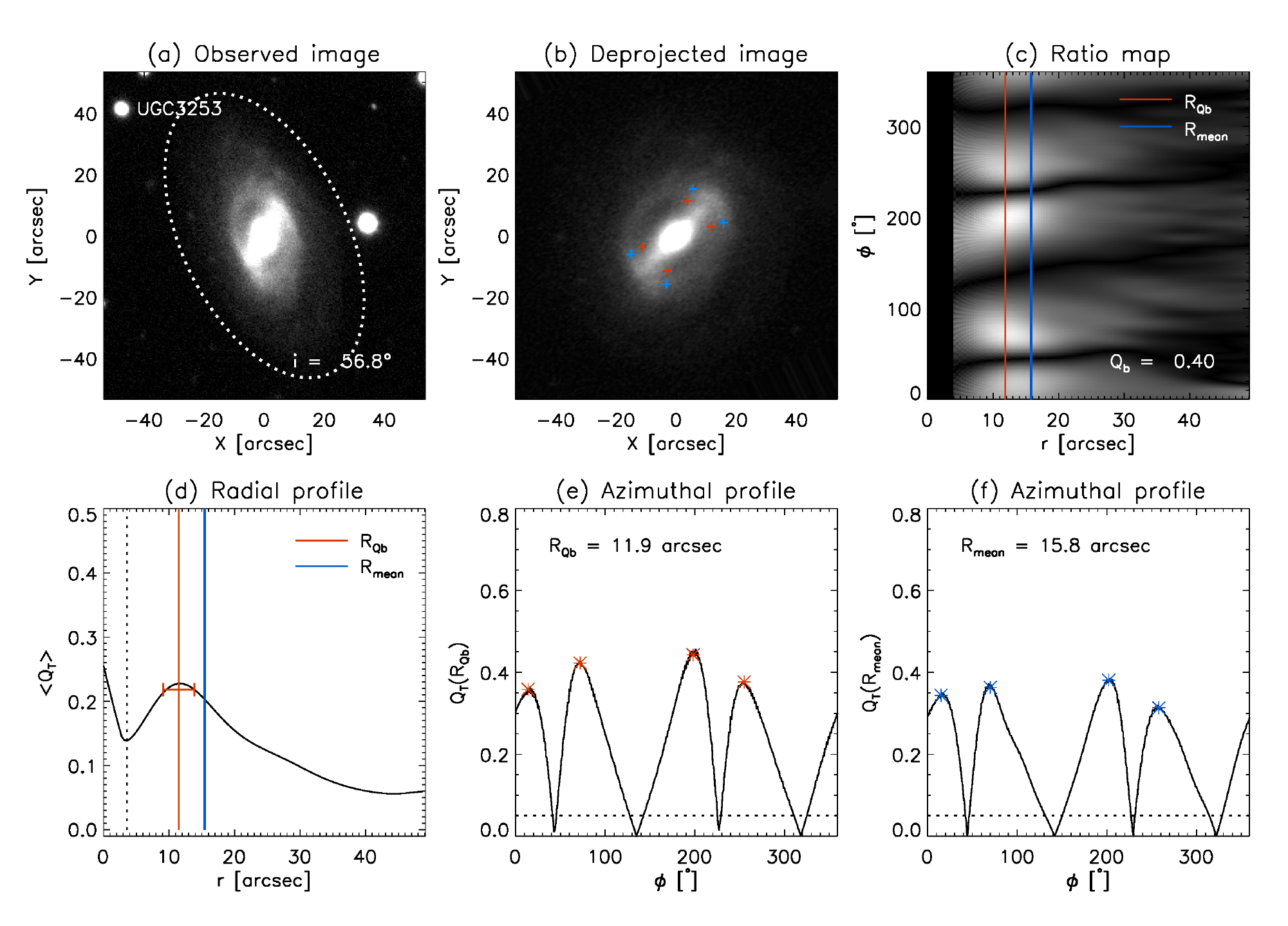}
    \caption{(continued).}
    \label{fig:pt_map6}
\end{figure*}

\section{Bar rotation rates with different bar radius measurements}

 The measurements of \rr\ obtained with all the available estimates of \rbar\ are presented in Table~\ref{tab:rotation_rates}.

\begin{table*}
\caption[\rr.]{Bar rotation rates obtained with different bar {\rm radii}.}
    \centering
    \begin{tabular}{ccccccc}
    \hline\hline
Galaxy & \rr$_{\rm 1}$ & \rr$_{\rm 2}$ & \rr$_{\rm 3}$ & \rr$_{\rm 4}$ & \rr$_{\rm 5}$ & \rr$_{\rm 6}$ \\
 (1) & (2) & (3) & (4) & (5) & (6) & (7)\\ 
\hline
IC~1528 & 1.13 & 0.74 & 0.58$^{+0.17}_{-0.20}$ & - & - & - \\
IC~1683 & 0.78 & 0.67 & 0.72$^{+0.35}_{-0.32}$ & 1.42 & 0.96 & 1.15\\
IC~5309 & 0.79 & 0.44 & 0.85$^{+0.73}_{-0.34}$ & 1.47 & 0.54 & - \\
NGC~36 & 0.61 & 0.51 & 0.83$^{+0.33}_{-0.27}$ & 0.68 & 0.56 & 1.08\\
NGC~2553 & 0.53 & 0.41 & 0.67$^{+0.14}_{-0.14}$ & 0.67 & 0.55 & 0.97\\
NGC~2880 & 1.04 & 0.92 & 0.50$^{+0.16}_{-0.15}$ & - & - & 0.89 \\
NGC~5205 & 0.76 & 0.61 & 0.57$^{+0.14}_{-0.13}$ & 0.73 & 0.61 & 0.80\\
NGC~5406 & 0.55 & 0.47 & 0.55$^{+0.21}_{-0.13}$ & 0.49 & 0.44 & 1.03\\
NGC~5947 & 0.60 & 0.46 & 0.54$^{+0.20}_{-0.26}$ & 0.46 & 0.46 & -\\
NGC~5971 & 0.91 & 0.30 & 1.03$^{+0.50}_{-0.46}$ & - & - & - \\
NGC~6497 & 0.44 & 0.35 & 0.34$^{+0.13}_{-0.11}$ & 0.27 & 0.27 & 0.52\\
UGC~3253 & 0.82 & 0.66 & 0.80$^{+0.21}_{-0.19}$ & 0.58 & 0.53 & 1.23\\
\hline
    \end{tabular}
    \\
\tablefoot{(1) Galaxy name. (2) Bar rotation rate obtained using $R_{\rm \epsilon,peak}$ as bar radius estimate. (3) Bar rotation rate obtained using $R_{\rm PA}$ as bar radius estimate. (4) Bar rotation rate obtained using $R_{\rm Fourier}$ as bar radius estimate. (5) Bar rotation rate obtained using $R_{\rm \epsilon}$ as bar radius estimate. (6) Bar rotation rate obtained using $R_{\rm trann}$ as bar radius estimate. (7) Bar rotation rate obtained using $R_{\rm A_2}$ as bar radius estimate.}
    \label{tab:rotation_rates}
\end{table*}

\end{appendix}

\label{lastpage}
\end{document}